\documentclass[nofootinbib,superscriptaddress,eqsecnum,tightenlines,11pt,longbibliography,notitlepage,twocolumn]{revtex4-1}

\usepackage[draft]{showlabels}
\usepackage{hyperref}
\usepackage{graphicx}
\usepackage{amsmath,amssymb,amsfonts,amsthm,stmaryrd,mathtools,mathtools,wasysym }
\usepackage{yfonts}
\usepackage{multirow,bigdelim}
\usepackage{dsfont}
\usepackage{color}
\usepackage{graphicx,color,epstopdf,epsfig,psfrag}
\usepackage[dvipsnames]{xcolor}
\usepackage{ mathrsfs }
\usepackage{footmisc}
\usepackage{enumerate}
\usepackage{tikz}
\usetikzlibrary{calc}
\usetikzlibrary{decorations.pathmorphing}
\usetikzlibrary{shapes.geometric}
\usetikzlibrary{arrows,decorations.markings}
\usepackage{bbold}

\hypersetup{
    colorlinks=true,       
    linkcolor=MidnightBlue,          
    citecolor=BrickRed,        
    filecolor=BrickRed,      
    urlcolor=BrickRed           
}
 
\makeatletter
\newcounter{subsubsubsection}[subsubsection]
\renewcommand\thesubsubsubsection{\thesubsubsection .\@alph\c@subsubsubsection}
\newcommand\subsubsubsection{\@startsection{subsubsubsection}{4}{\z@}%
                                     {-3.25ex\@plus -1ex \@minus -.2ex}%
                                     {1.5ex \@plus .2ex}%
                                     {\centering\normalfont\small\textit}}
\newcommand*\l@subsubsubsection{\@dottedtocline{3}{10.0em}{4.1em}}
\newcommand*{\subsubsubsectionmark}[1]{}
\makeatother

\numberwithin{equation}{section}
\numberwithin{paragraph}{subsection}

\renewcommand{\Re}{{\mathrm{Re}}}

\newcommand{\cH}{{\mathcal H}}

\newcommand{\cP}{{\mathcal P}}

\newcommand{\cD}{{\mathcal D}}

\newcommand{\SU}{\mathrm{SU}}
\newcommand{\ISU}{\mathrm{ISU}}

\newcommand{\SL}{\mathrm{SL}}
\newcommand{\SO}{\mathrm{SO}}

\renewcommand{\d}{{\mathrm{d}}}
\newcommand{\D}{{\mathrm{D}}}
\newcommand{\dd}{{\delta}}

\newcommand{\EOMeq}{{\,\stackrel{\cdot}{=}\,}}

\newcommand{\be}{\begin{equation}}
\newcommand{\ee}{\end{equation}}
\newcommand{\beq}{\begin{eqnarray}}
\newcommand{\eeq}{\end{eqnarray}}
\newcommand{\bes}{\begin{eqnarray}}
\newcommand{\ees}{\end{eqnarray}}

\newcommand{\mat} [2] {\left ( \begin{array}{#1}#2\end{array} \right ) }

\newcommand{\su}{{\mathfrak{su}}}

\newcommand{\ad}{{\mathrm{ad}}}
\newcommand{\Ad}{{\mathrm{Ad}}}

\newcommand{\la}{\langle}
\newcommand{\ra}{\rangle}

\newcommand{\Tr}{{\mathrm{Tr}}}
\newcommand{\tr}{{\mathrm{tr}}}
\newcommand{\f}{\frac}

\def\pp{\partial}

\def\eps{\epsilon}

\newcommand{\h}{{\mathrm{h}}}
\newcommand{\G}{{\Gamma}}

\renewcommand{\hat}{\widehat}

\renewcommand{\bar}{\overline}

\newcommand{\ellpl}{\ell_\text{Pl}}

\theoremstyle{definition}

\theoremstyle{remark}

\begin{document}

\title{\LARGE Quantum edge modes in 3d gravity\\ and 2+1d topological phases of matter }

\author{{\bf Aldo Riello}}\email{ariello@perimeterinstitute.ca}
\affiliation{Perimeter Institute, 31 Caroline St North, Waterloo ON, Canada N2L 2Y5}

\date{\today}

\begin{abstract}
We analyze the edge mode structure of Euclidean three dimensional gravity from within the quantum theory as embodied by a Ponzano--Regge--Turaev--Viro discrete state sum with Gibbons--Hawking--York boundary conditions.
This structure is encoded in a pair of dual statistical models of the vertex and face kind, which for specific choices of boundary conditions turn out to be integrable.
The duality is just the manifestation of a pervasive dual structure which manifests at different levels of the classical and quantum theories.
Emphasis will be put on the geometrical interpretation of the edge modes which leads in particular to the identification of the quantum analogue of Carlip's would-be normal diffeomorphisms.
We also provide a reinterpretation of our construction in terms of a non-Abelian 2+1 topological phase with electric boundary conditions.
\end{abstract}

\maketitle


\section{Introduction}

{\bf Edge modes}
Consideration of physical theories in finite bounded regions entails a choice of boundary conditions that generically requires breaking the gauge invariance  at the region's boundary.
Gauge invariance can be fully restored by the introduction of compensating fields at the boundary, often referred to as `edge modes'.
As a consequence of the original gauge invariance, the edge theory has a large symmetry group%
\footnote{Symmetries distinguish themselves from gauge invariances because they have associated non-vanishing charges.}
(e.g. \cite{Elitzur1989, Wen1992, Balachandran1992, Balachandran1996, ArcioniBlau2002, Carlip2005b, DonnellyFreidel2016}.)

To discuss the physicality of the edge modes, it is useful to introduce a distinction between `physical' and `abstract' boundaries. 
By `physical' we indicate the actual edge of a chunk of metal or the interface between two materials; while by `abstract' we mean purely theoretical subdivisions of a region in two adjacent subregions.%
\footnotetext{Physically, boundaries are always interfaces, either between two materials or between regions of space(time). It is useful to keep in mind this simple observation, especially when referring to the `bounding surface' as a seemingly independent entity.}

In the case of physical boundaries, preservation of gauge invariance---be it effective as in quantum Hall states, or fundamental as in electrodynamics---and avoidance of anomalies require gauge fields to couple to { something} which does actually live on the bounding surface, e.g. some electric charge density.
In this sense, the { full} system does not require the introduction of any { new} edge mode.
This is true in particular when the boundary represents the interaction surface between the system and a measuring apparatus.

The case of abstract boundaries is hence most easily understood as an idealization of the first case, and the introduction of abstract compensating fields as the simplest { model} of a physical boundary or measuring device. 
To achieve this, a simple possibility is that the compensating fields coordinatize the fibers of the principal fiber bundle on which the gauge theory is constructed---directions that explicitly manifest themselves only at the boundaries because of gauge invariance itself.%
\footnote{On a boundary chart $\partial U$, the coordinates on the principal fiber bundles are $(y, g)\in \partial U\times G$. Fixing the gauge at the boundary defines the function $g(y)$, that can hence be promoted to be the `compensating' field. Compensating fields do not necessarily have to be of this form.}
(A more general, and yet more minimal, setup is discussed in \cite{GomesRiello2017} under the name of `field space connection'. This framework helps modeling cases where an `abstract' measuring device---or observer---is composed of physical fields present also in the bulk, a scenario particularly relevant for a theory of gravitation.)

{\bf A gauge theoretical example}
The prototypical example of the principal fiber bundle construction is the derivation of the Wess--Zumino--Witten model from the 3d Chern--Simons action \cite{Elitzur1989, Wen1992, Balachandran1992, Balachandran1996, ArcioniBlau2002, Carlip2005b}.
The Chern--Simons action one has to start from features a boundary contribution that guarantees, even in presence of boundaries, its functional differentiability with respect to the pullback of the connection on the boundary along one of its two intrinsic directions. In formulas,\footnote{We omit the normalized coupling constant $k/4\pi$, $k\in\mathbb Z$, in front of the action.} 
\be
\text{CS}_z[A]= \int \Tr[ A\wedge\d A + \tfrac23 A^{\wedge 3}] -i \oint A_z A_{\bar z}
\ee
so that ($\EOMeq$ indicates equality on-shell)
\be
\dd \text{CS}_z[A] \EOMeq  - 2i \oint \Tr[ A_{\bar z}\dd A_z ].
\ee
The Wess--Zumino--Witten model then arises from the comparison of the Chern--Simons action evaluated on two different, gauge related, configurations $A$ and $A^G = G^{-1} (A + \d) G$:%
\footnote{The bulk term in the Wess--Zumino--Witten action is crucial for the quantum consistency of the theory \cite{Witten1984}, but classically does not play any dynamical role.}
\begin{subequations}
\begin{align}
\text{WZW}&[G|A_z] = \text{CS}_z[A] - \text{CS}_z[A^G] \\
 =& i\oint \Tr[\pp_z GG^{-1} \pp_{\bar z} GG^{-1} + 2A_z \pp_{\bar z} GG^{-1}]  \notag\\
 &+\tfrac13\int \Tr[ (\d GG^{-1})^{\wedge 3}].
\end{align}
\label{eq_WZW}
\end{subequations}
The resulting compensating fields are hence valued in the gauge group, representing the local `gauge frames' at the boundary, while from the edge mode perspective the boundary value of $A_z$ is a background (classical) field.

The local `gauge frame' $G$ is akin to the local Lorentz frame, or maybe 3d orientation, of a fleet of a particle detector: it does {not} have any absolute meaning but it is still necessary to fix it {somehow} in order to successfully compare particle momenta.
This is especially needed when the particles reach the detector from two different sides of the bounding surface. 
A symmetry group acting on the edge modes simply reflects the freedom in the fixing of the detector's orientation. 
The difference with gauge invariance is subtle and spurs solely from the demands of an eventual gluing of the two regions.
From the principal fiber bundle perspective, this corresponds to the need of gluing consistently not only the base manifolds but the whole bundles.\footnote{Of course, {\it known} transition functions can be used in the gluing procedure.}

{\bf Diffeomorphisms}
Among theories with local symmetries, general relativity has a somewhat special status.
This is because its local symmetry is diffeomorphism invariance.
The latter can be seen as acting either actively, by displacing the fields on the spacetime manifold, or passively, by relabeling the points of the spacetime manifold.
The so-called `hole argument' shows that this symmetry implies that spacetime points have no physical meaning per se, i.e. in absence of fields, and events can only be localized with respect to each other, rather than with respect to the underlying manifold \cite{Einstein1916, Rovelli2004}.

Similarly to gauge theories, therefore, also in general relativity physical boundaries are defined by the presence of `something'.
Differently from gauge theories, however, one cannot suppose that such boundaries have a fixed location, or that the matter fields defining the boundary are non-dynamical, because either condition would fundamentally violate diffeomorphism invariance.
Hence, one must appeal to a relational definition of the boundary surface, e.g. as the level surface of some dynamical scalar quantity.
Again, for physically (relationally) defined boundaries there is no need to introduce compensating fields. 

Given the difficulties of working in a fully relational approach, also in general relativity it is useful to study compensating fields analogous to the above.
At this purpose, one can introduce a set of `preferred' near-boundary coordinates morally representing a network of spacetime beacons.

The striking physical content of general covariance---sometimes obscured by the sheer power of the geometrical formalism---is that {\it any} such network of beacons can be used as a viable reference system and predictions are independent of this choice.
Now, with the idea that boundaries are about the relation between a system and a measuring apparatus, it is clear that boundaries must know about the beacon system.

To detect the edge modes, it is enough to proceed as in the Chern--Simons - Wess--Zumino--Witten case. 
First, we pick an action which is differentiable, e.g. with respect to the induced metric, even in presence of boundaries.
This requires the Einstein--Hilbert action to be augmented by the York--Gibbons--Hawking boundary term \cite{GibbonsHawking,York1986}.
We introduce then a diffeomorphism-breaking beacon system to fix the position of the boundary, and evaluate the gravitational action on two diffeomorphism related configurations. 
Due to the relation between active and passive diffeomorphisms, it is enough to consider displacements of the boundary.
With notation adapted to 3 spacetime dimensions, $\ellpl = 8\pi G_\text{N}$ ($\hbar=1$), and a boundary set at the value $\rho$ of a `radial' coordinate $r$,\footnote{As usual: $g_{\mu\nu}$ is the three-metric on $M$, $R$ its Ricci scalar, $\Lambda$ the cosmological constant, $h_{\mu\nu}$ the induced metric on $\pp M$, and $K$ the trace of the extrinsic curvature of $\pp M$. Finally, $g=|\det g_{\mu\nu}|$ and similarly for $h$. }
\be
\text{GR}_\rho[g_{\mu\nu}] = \f{1}{2\ellpl}\left[\int\sqrt{g}  (R - 2\Lambda) +2\oint_{r=\rho} \sqrt{h} K\right]
\label{eq_GRact}
\ee
so that 
\be
\dd \text{GR}_\rho[g_{\mu\nu}] \EOMeq \frac{1}{\ellpl} \oint \sqrt{h}(K^{\mu\nu}-Kh^{\mu\nu})\dd h_{\mu\nu} .
\label{eq_deltaGRact}
\ee
For an infinitesimal displacement $\varphi$, the edge mode action is schematically
\be
S_\text{edge}[\varphi|h_{\mu\nu}] = \text{GR}_\rho[g] - \text{GR}_{\rho+\varphi}[g],
\label{eq_edgeGR}
\ee
where $h_{\mu\nu}$ serves as a background metric for the edge mode field $\varphi$.

In the case the boundary beacon system is left untouched---i.e. if a gauge fixing of the boundary coordinate system is chosen---the only remaining compensating field will be the one associated to radial diffeomorphisms.
This is the core of Carlip's derivation of the Liouville field theory as a `would-be gauge' edge mode at the conformal boundary of an asymptotically AdS$_3$ spacetime \cite{Carlip2005a} (see also \cite{Carlip2005b} and references therein).%
\footnote{Excluding singularities, asymptotitc infinity, $\mathscr{I}$, is possibly the only example of an actual boundary in general relativity, at least when considered in the `unphysical', i.e. conformally rescaled, spacetime. From the viewpoint of the physical spacetime, $\mathscr{I}$ exists only as an idealization of a `far away' surface defined as the limit of a set of level surfaces of an (auxiliary) scalar field, which plays the role of a radial coordinate in the above sense. }

{\bf Goal and layout of the article}
The goal of this article is to give a quantum mechanical account of the edge mode theories of Euclidean three dimensional gravity, and in particular of the quantum analogue of the field $\varphi$ above.
The geometrical picture is most transparent in absence of a cosmological constant and in the covariant picture, and for this reason we will mostly concentrate on this case.
The positive cosmological case, as well as the canonical formulation, will be also touched upon and will allow to draw a link with the theory of topological phases of matter.

We start by reviewing first order gravity as a $BF$ topological gauge theory, Sec. \ref{sec_BF}.
Emphasis will be put on the structure of its symmetries and its quantization will be sketched in both connection and metric variables,  Sec.s \ref{sec_quant} and \ref{sec_metricbc} respectively. The latter will lead us to consider the Ponzano--Regge--Turaev--Viro state sum.
After a brief discussion of the bulk symmetry of the model, Sec. \ref{sec_bulksym}, we move to the core of the paper. In Section \ref{sec_edgemodes}, we present the quantum edge modes, with a focus on the pair of dual theories emerging from the symmetry structure of $BF$ theory.
The so far local analysis is then integrated with information on how to deal with handlebody topologies from a purely boundary perspective, Sec. \ref{sec_torus}.
At this point we exemplify the proposed constructions and dualities with the explicit example of an integrable edge theory, Sec. \ref{sec_onehalf}.
A graphical notation is introduced in Sec. \ref{sec_graph}, which will help us to deal more efficiently with the last part of this article concerning the case of a non-vanishing cosmological constant, Sec. \ref{sec_cc}, as well as with the translation to the canonical picture and the theory of topological phases of matter, Sec. \ref{sec_canonical}. 
We summarize and conclude in Sec. \ref{sec_conclusions}.
The article finally contains two appendices, one  of which---App. \ref{app_triadbc}---is devoted to considerations about fixed-triad boundary conditions in the first-order formulation of gravity.

The accustomed reader might want to refer to the Sec. \ref{sec_graph} to decode some of the equations of Sec.s \ref{sec_metricbc} and \ref{sec_edgemodes}.

\section{First oder gravity\label{sec_BF}}

In absence of a cosmological constant, $\Lambda=0$, the action of Euclidean 3d gravity in the first-order formalism is that of an $\ISU(2)$ $BF$ topological field theory \cite{Witten1988,Horowitz1989,Matschull1999}: 
\be
S_\omega[\omega,e] = \f{1}{\ellpl}\int_M \la e\stackrel{\wedge}{,} F[\omega] \ra,
\label{eq_Somega}
\ee
where the conjugated variables $\omega=\omega^a_\mu \d x^\mu J_a$ and $e=e^a_\mu \d x^\mu P_a$ are the spin connection and local frame field (triad), respectively.
Here, $(J_a,P_a)$ is a basis of the Lie algebra
\be
\mathfrak{g} =\mathfrak{isu}(2)\cong\su(2)\ltimes_\ad \su(2)_+ \ni(J_a,P_a)
\label{eq_isu}
\ee
where $\su(2)_+$ stands for $\su(2)$ seen as an Abelian group (isomorphic to its own Lie algebra) under the addition operation:
\be
[J_a,J_b] = \eps_{ab}{}^c J_c,
\;\,
[J_a,P_b] = \eps_{ab}{}^c P_c,
\;\,
[P_a,P_b]=0.
\ee
The symbol $F$ stands for the curvature of $\omega$, $F = \d \omega +[ \omega\stackrel{\wedge}{,}\omega]$, while $\la\cdot,\cdot\ra$ 
is the bilinear form (this symbol will be left understood in the following):%
\footnote{Using the second isomorphism of Eq. \eqref{eq_isu}, one can re-express $S_\omega$ as an $\SU(2)$ $BF$ theory, by setting $J_a = -\frac{i}{2}\sigma_a = P_a $ and $\la\cdot,\cdot\ra = 2\Tr( \cdot \cdot)$ with the trace taken in the fundamental representation of $\su(2)$. }
\be
\la J_a, P_b\ra = \delta_{ab},
\quad
\la J_a,J_b\ra = 0 = \la P_a,P_b\ra.
\ee 

The (on-shell) relation of $e$ and $\omega$ to the spacetime metric and Christoffel symbol is
\be
g_{\mu\nu} = \delta_{ab} e^a_\mu e^b_\nu,\;\;
\Gamma^\mu_{\rho\sigma} = (e^{-1})^\mu_a \left(\pp_\rho e^a_\sigma + \epsilon^a{}_{bc}\omega^b_\rho e^c_\sigma\right).
\ee

The subindex `$\omega$' in $S_\omega$ indicates that in presence of boundaries, $\pp M\neq \emptyset$, this action is differentiable with respect to the connection variable:
\be
\dd S_\omega = \f1\ellpl\int_M \dd e \wedge F + D_\omega e \wedge \dd \omega  - \f1\ellpl \oint_{\pp M}  e\wedge \dd \omega.
\label{eq_dSomega}
\ee
Notice that the equations of motion imply the flatness of $\omega$ and torsion-freeness of $e$:
\be
F \EOMeq 0,
\quad
\D_\omega e \EOMeq 0.
\ee
General (local) solutions to these equations are, 
\be
\omega = g^{-1} \d g,
\quad
e = g^{-1} \d \lambda g.
\ee
As we will review in a moment, these expressions indicate that $\omega$ and $e$ are `pure-gauge', which  testifies the topological nature of 3d gravity. 

The (bulk) invariances of $S_\omega$ are generated by infinitesimal ($i$) spacetime (passive) {\it diffeomorphisms},\footnote{According to the Cartan formula, the Lie derivative acts on one-forms $\eta\in\Omega^1(M)$ as $\pounds_\xi\eta = \d i_\xi \eta + i_\xi \d \eta$.} $\xi\in\mathfrak{X}^1(M)$
\be
x\mapsto x + \xi,
\quad
\delta_\xi e = \pounds_\xi e, 
\quad
\delta_\xi \omega = \pounds_\xi \omega; 
\label{eq_diff}
\ee
($ii$) local {\it Lorentz} (gauge) symmetry\footnote{The term `symmetry' here is used in a looser sense than in the introduction, meaning a transformation that leaves the action invariance.}, $X\in {\cal C}^\infty(M,\su(2))$
\be
\delta_X e = -\ad_X e,
\quad
\delta_X \omega = -\ad_X \omega + \d X = \D_\omega X;
\ee
and ($iii$) {\it shift} symmetry, $\lambda\in{\cal C}^\infty(M,\su(2)_+)$
\be
\delta_\lambda e = e + \D_\omega \lambda,
\quad
\delta_\lambda \omega = 0,
\label{eq_flatshift}
\ee

The internal symmetries are nicely compatible with each other, and locally organize themselves in the Poisson--Lie group structure
\be
G = \mathrm{ISU}(2) \cong \SU(2) \ltimes_\Ad \su(2)_+ \cong \mathrm{T}^*\SU(2).
\label{eq_ISO}
\ee 
The first isomorphism emphasizes the Lie structure, while the second the Poisson structure of $G$. The two satisfy compatibility requirements.
This Possoin--Lie symmetry is the classical counterpart of a quantum double (Hopf algebra) symmetry of the quantum theory.

It is useful to think of $\SU(2)$ Lorentz symmetry as being associated to the connection variable $\omega$, while $\su(2)_+$ shift symmetry to the triad $e$.

The origin of shift symmetry lies in the Bianchi identity $\D_\omega F \equiv 0$.
Shift symmetry is peculiar to 3d gravity and is the symmetry that makes this theory topological and solvable.
On a flat background (i.e. on-shell of the equation of motion), shift symmetry plays the role of {active diffeomorphisms}.
A first hint of this fact is that (on-shell) the action of an infinitesimal diffeomorphism on the fields is reproduced by a field-dependent shift transformation $\lambda = i_\xi e$ (modulo a field-dependent Lorentz transformation $X=i_\xi \omega$) \cite{FreidelLouapre2003,BaratinGirelliOriti2011}.
Spacetime points are left untouched.

From a canonical perspective, on a spacetime of the form%
\footnote{We ignore here the Lorentz structure of the spacetime $M$. First, because we are dealing with Euclidean gravity, second because if we were dealing with a Lorentzian theory, the Lorentz structure would be fixed only dynamically by a given configuration of the triad field.}
$M\cong\Sigma\times \mathbb R$, $\pp\Sigma=\emptyset$, the conjugated variables on phase space are 
\be
A = \underleftarrow \omega,
\quad
E = \underleftarrow e,
\ee
where the under-arrow stands for the pullback onto $\Sigma$, with 
\be
\Big\{ E^a_\mu(x) , A^b_\nu(y) \Big\} = \delta^{ab}\eps_{\mu\nu}\delta^{(2)}(x-y)
\ee
(all other brackets vanish).

The remaining components of $e$ and $\omega$ are Lagrange multipliers for the first class constraints
\be
C_\text{L} = \underleftarrow{\D_\omega e} = \D_A E,
\quad
C_\text{sh} = \underleftarrow {F[\omega]} = F[A],
\label{eq_constr}
\ee
which symplectically generate on $(A,E)$ Lorentz and shift symmetry, respectively. They are known as the Gauss and flatness constraint.

In presence of boundaries, the action $S_\omega$ is fully invariant under Lorentz transformations, and changes by a boundary term proportional to $F$ under a shift symmetry transformations:
\be
\delta_X S_\omega =0, 
\quad
\delta_\lambda S_\omega = \f1\ellpl \oint_{\pp M} \lambda F \EOMeq 0.
\label{eq_deltaSomega}
\ee
This makes the connection polarization particularly well suited for quantization. (See App. \ref{app_triadbc} for a brief discussion of fixed-triad boundary conditions.)

\section{Quantization \label{sec_quant}}

We will now proceed formally to sketching the quantization of this theory.
Its justification can be found either in the more rigorous treatments of \cite{NouiPerez2005, FreidelLouapre2004, FreidelLouapre2004b, BarrettNaishGuzman2009, MeusburgerNoui2010,BonzomSmerlak2012}, or in the fact that a direct deformation of the final result leads to a well-defined topological field theory, the $U_q(\SU(2))$ Turaev--Viro model, which is equivalent to the quantization of Chern--Simons gravity for $\Lambda>0$ \cite{Turaev1990} (see Sec. \ref{sec_cc}). In any case, we will highlight various hints to the correctness of our manipulations.

In the Schroedinger picture, define
\be
Z_M[A] = \int_{A} \cD \omega \cD e \, e^{i S_\omega[\omega,e]} = \int_{A} \cD \omega \delta(F[\omega]),
\ee
where the subscript $A$ means that the functional integral is performed keeping $\underleftarrow \omega=A$ fixed.
The above formula says that $Z_M[A]$ is a uniform distribution on the moduli space of boundary-flat connections $A$ admitting a bulk-flat extensions $\omega$.
Notice the dependence on the topology of $M$.

To help make sense of this expression, introduce a discretization $\Delta$ of $M$, such that $\pp \Delta$ is a discretization of $\partial M$.
$\Delta$ is a cellular complex, whose $d$-dimensional cells $c_d$ are denoted for growing $d=0,\dots,3$, vertices $v$, edges $l$, faces $f$, and 3-cells $\sigma$, respectively.
It is conventional to discretize the connection along the Poincar\'e dual complex $\Delta^\ast$. 
The treatment is consistent if we assign dual cells separately in the bulk and on the boundary of $\Delta$: in the bulk we set $c_d^\ast \leftrightarrow c_{3-d}$, while on the boundary $c_d^\ast \leftrightarrow c_{2-d}$. In other terms, we demand for the boundary dual graph $\Gamma$, that
\be
\Gamma := \pp \Delta^\ast = (\pp \Delta)^\ast.
\ee

In particular, consider a (directed) dual edge $l^\ast$ extending between source and target dual vertices, $s(l^*)$ and $t(l^*)$, and define along it the parallel transport or, with a slightly improper language, the holonomy
\be
h_{l^*} = P\exp \int_{l^\ast} \omega.
\ee
Notice that in the bulk $l^\ast$ is a dual edge which crosses a face $f$ shared between two adjacent 3-cells $\sigma_1$ and $\sigma_2$, while on the boundary $l^\ast$ is a dual edge which crosses a boundary edge $l_\pp$ shared between two adjacent boundary faces $f_{\pp,1}$ and $f_{\pp,2}$.

Gauge transformations act at dual vertices,
\be
h_{l^*} \mapsto g_{t(l^*)} h_{l*} g_{s(l^*)}^{-1},
\label{eq_gtr}
\ee 
while flatness means that along each dual face the holonomy is trivial,
\be
H_{f^*}=\overleftarrow{\prod_{l^\ast\in\pp f^\ast}} h_{l^\ast}^{\eps(l^\ast,f^\ast)} = \mathbb 1 .
\label{eq_Hfast}
\ee
Here $\eps(l^\ast,f^\ast)=\pm1$ according to the relative orientation of $f^\ast$ and $l^\ast$; to avoid clutter, we leave it understood in the following formulas. 

Shift transformations have a trivial action in the holonomy polarization.

Supposing that $\cD\omega$ is a uniform measure, we discretize $Z_M[A]$ as
\be
Z_\Delta[h_{l^\ast_\pp}] = \Big[ \prod_{l^\ast\notin\pp\Delta^\ast}\int \d h_{l^\ast} \Big] \prod'_{f^\ast\in \text{Int}(\Delta^*)}\delta(H_{f^*}),
\label{eq_ZDelta}
\ee
where $\d h$ is the Haar measure on $\SU(2)$, $\text{Int}(\Delta^*)$ stands for the bulk part of the dual discretization, and the prime on the product means that certain faces are omitted to avoid redundancies among delta functions. More generally, they can be replaced with any class function of the total face holonomy with value 1 at the identity. 
This makes $Z_\Delta$ into a well-defined distribution over%
\footnote{See \cite{BonzomSmerlak2012,BonzomSmerlak2012a} for possible subtleties on more involved topologies.}
$\mathcal H'_\Gamma = L^2\left(\SU(2)^{\times L^\ast}\right)$, $L^\ast=\#\{l^\ast\in\Gamma\}$. 
This is the gauge-variant Hilbert space of discretized connections over $\Gamma$.

Because of  Eq. \eqref{eq_deltaSomega}, $Z_M[A]$ is (on-shell) formally invariant both under Lorentz and shift transformations. And so is $Z_\Delta$.
In particular, it acts as a projector over the gauge invariant part of $\mathcal H'$, to which we will now restrict:
\be
\mathcal H_\Gamma = L^2\left(\SU(2)^{\times L^\ast}//\SU(2)^{\times V^*}\right),
\ee
where $V^\ast=\#\{v^\ast\in\Gamma\}$.

\section{Metric boundary conditions and the Ponzano--Regge model \label{sec_metricbc}}

So far we have worked with boundary conditions that require the boundary connection $A$ to be fixed.
This was so because we started from a path integral formulation based on an action differentiable in $A$, i.e. $\dd S_\omega \EOMeq -\ellpl^{-1}\oint E\wedge \dd A$.
Equation \eqref{eq_deltaSomega} explains why this action principle is a convenient choice: it is fully compatible with the symmetries.

What if we wanted to consider more general boundary conditions? 
In particular, what if we wanted to fix the induced boundary metric in a quantum analogue of Eq.s \eqref{eq_GRact} and \eqref{eq_deltaGRact}?

Building a quantization from an action differentiable in $E$ is possible, but various difficulties have to be overcome.
The reason is that such an action, $S_e = S_\omega + \ellpl^{-1}\oint e\wedge\omega$, is neither Lorentz nor shift invariant (even on-shell of the constraints).
And, in this form, it does not admit a fully natural discretization either---see, however, \cite{DupuisFreidelGirelli2017} and also App. \ref{app_triadbc}.

An alternative perspective consists in looking within ${\cal H}_\Gamma$ for superpositions of boundary connections which fix some property of our interests. 
In other words, `boundary states' $\Psi[A]\in{\cal H}_\Gamma$ can be thoughts as implementing different boundary conditions to the path integral.

In other words, the expression
\be
\la Z_\Delta | \Psi_o \ra = \Big[ \prod_{l^\ast\in\pp\Delta^\ast}\int \d h_{l^\ast} \Big] { Z_\Delta[h_{l^*_\pp}] }\Psi_o[h_{l^*_\pp}]
\label{eq_ampl}
\ee
can be thought as the implementation of the path integral
\be
\int_{\psi=\Psi_o} \cD\omega\cD e\, e^{i S_\psi}
\label{eq_heuristic}
\ee
where $\psi$ represents a class of boundary conditions and $\Psi_o$ a particular choice therein, while $S_\psi = S_\omega + (\text{\it some bdry term})$ is the corresponding action principle.

Let us focus on metric, or York--Gibbons--Hawking, boundary conditions. 
In the discrete context we expect these to correspond to a state diagonalizing the lengths of the boundary edges $l_\pp\in\pp\Delta$.
Constructing this kind of states, known as {\it spin-networks}, was one of the early successes of loop quantum gravity \cite{RovelliSmolin1995}.

They read
\be
\Psi_{(j,\iota)}[h_{l^\ast_\pp}] =\tr_\Gamma \Big[ \bigotimes_{l^\ast_\pp\in\Gamma} D^{j_l}(h_{l^\ast_\pp})  \bigotimes_{v^\ast_\pp\in\Gamma} \iota_{v^\ast_\pp} \Big].
\label{eq_SN}
\ee
Here, $D^j(h): V_j\to V_j$ is a Wigner matrix in the spin-$j$ representation of $\SU(2)$, and $\iota_{v^\ast} \in \text{Inv}\Big( \bigotimes_{l^\ast:v^\ast\in\pp l^\ast}  V_{j_{l\ast}} 
\Big)$ is an intertwining operator associated to the (original) dual vertices of $\Gamma$. It implements gauge invariance. In our conventions every dual vertex is outwardly oriented. Supposing it is $N$ valent:
\be
\Big(\bigotimes_{i=1}^N D^{j_i}(g)^{n_i}{}_{m_i} \Big) \iota^{m_1\cdots m_N} = \iota^{n_1\cdots n_N}.
\ee
Dual edges are oriented so that in the matrix element $D^{j}(h)^{n}{}_{m}$ the indices $m$ and $n$ are attached to the source and target vertices, respectively.
Finally, $\tr_\Gamma$ represents the contraction of all the magnetic indices according to the pattern determined by the dual boundary graph $\Gamma$, leaving understood that two upper indices are contracted via the spin $j$ generalization of the $\SU(2)$ invariant tensor $\epsilon_{n'n}=\pm1$, i.e. $(-1)^n\delta_{n',-n}$, which intertwines between $V_j$ and its contragradient representation $\bar V_j$ (thus adjusting for discording orientations at the targets of $l^*$). See \cite{BarrettNaishGuzman2009} for details.
This construction guarantees gauge invariance, i.e.
\be
\Psi_{(j,\iota)}[h_{l^\ast_\pp}] = \Psi_{(j,\iota)}[ g_{t(l^*_\pp)} h_{l*_\pp} g_{s(l^*_\pp)}^{-1}]
\label{eq_ginv}
\ee
for any choice of $\{g_{v^*_\pp}\in\SU(2)\}$ (cf. Eq. \eqref{eq_gtr}).

Notice that for $\Delta$ a triangulation, the dual vertices are trivalent and the intertwiners unique and equal to $3j$ symbols---i.e., modulo dualizations, to Clebsh--Gordan coefficients.

Here, the spin $j_{l}\in\f12\mathbb N$ is the quantum number associated to the length operator along $l$.
This operator corresponds to the quantization of 
\be
\ell_{l} = \left|\left| \int_0^1 \Ad_{h(\tau)}e_\mu(l(\tau)) \f{\d l^\mu}{\d \tau} \d \tau \right|\right|,
\label{eq_length}
\ee
where $h(\tau)$ is the $\omega$-parallel transport along $l$ from $s(l)=l(\tau=0)$ to $l(\tau)$, and the norm in $\su(2)$ is defined by $||X|| = \sqrt{\delta_{ab} X^a X^b}$. Notice that for flat and torsionless configurations, by Stokes theorem, the `length' of a closed loop vanishes. 
Thus, on shell, $\ell_l$ has to be interpreted as the {geodesic distance} between the endpoints of the edge $l$, rather than the length of a curve.
Its spectrum is given by the value of the Casimir \cite{Rovelli1993, FreidelLivineRovelli2003},
\be
\sqrt{j(j+1)}.
\ee

Following the Peter--Weyl theorem, $D^j(h)$ is best understood as the non-Abelian generalization of the Fourier transform providing the spectral decomposition of square integrable functions on (multiple copies of) $\SU(2)$.
What is suggested by this construction is that the analogy with the Fourier transform goes further to include the property of transforming one polarization of the quantum wave function to its conjugate one.

In turn, this suggests to apply the transform to $Z_\Delta$ itself, so to obtain a purely metric formulation of 3d quantum gravity.
In the case of a triangulation, this is well known to lead to the Ponzano--Regge model \cite{BarrettNaishGuzman2009}:
\begin{align}
&\hat Z_\Delta[j_{l_\pp}] 
= \la Z_\Delta | \Psi_{(j,\iota)}\ra\label{eq_PR}\\
&= \sum'_{\{j_l : l\notin\pp\Delta\}}\prod_l (-1)^{2j_l}d_{j_l} \prod_f (-1)^{k_f} \prod_\sigma \{6j\}_\sigma\notag
\end{align}
where $d_j  = \text{dim}(V_j)= 2j+1$,  $k_f = \sum_{l\in\pp f} j_l$ and $\{6j\}_\sigma$ is a $6j$ symbol associated to the lengths of the sides of a tetrahedron.
The prime means that one keeps fixed those spins attached to those edges which correspond to the omitted delta functions in \eqref{eq_ZDelta}.

The above formula is essentially a consequence of the following two identities:\footnote{Recall that a $6j$-symbol is essentially a contraction of four Clebsch--Gordan coefficients.}
\begin{align}
&\delta(h) = \sum_{j} d_j \chi^j(h)  \label{eq_delta}\\
&\int \d g \big(\overline{D^J}\otimes D^{j} \otimes D^{j'}\Big)(g) = \delta_{Jjj'} C^J_{jj'}\otimes \overline{C^J_{jj'}} \label{eq_Clebsch}
\end{align}
where $\chi^j(h) = \tr(D^j(h))$ is the spin $j$ character, $\delta_{Jjj'}$ is 1 if the three spins satisfy the triangular inequalities and zero otherwise, and $(C^J_{jj'})^M_{mm'}=\sqrt{d_J}\la JM|j j', mm'\ra$ is a rescaled Clebsch--Gordan coefficient.

What prompted Regge and Ponzano to propose the above as a quantum gravitational model of 3d gravity in 1968 is the fact that for large quantum number (homogeneously large spins), the asymptotic of the $6j$ symbol reproduces a discretized version of the Einstein--Hilbert--York--Gibbons--Hawking action, proposed a few year before by Regge himself \cite{PonzanoRegge1968,Roberts1999,HaggardLittlejohnEtAl2012}.

Even more compelling evidence emerges from the fact that the Biedenharn--Elliot identity for the $6j$-symbols admits the interpretation of a discrete version of the action of the Hamiltonian constraint, and other related facts \cite{BarrettCrane1997, BonzomFreidel2011, BonzomDittrich2013}.

\section{Bulk symmetries\label{sec_bulksym}}

As we have already observed, in the connection representation, shift transformations act trivially, while invariance under Lorentz transformations is ensured by the structure of the amplitude.

In the dual Ponzano--Regge formulation, on the other hand, Lorentz transformations act trivially.
This is because the utilized variables are the lengths $\ell_l$ rather than the triads $e^a$.
To track the action of shift symmetry, we can look at Eq. \eqref{eq_length}: on a flat background, the geodesic distances between the endpoints of a path $l$ associated to two shift-related triads, say $e$ and $e+\D_\omega\lambda$, differ exactly by $||\Delta \lambda|| = ||\lambda(t(l)) - \lambda(s(l))||$.
This observation justifies the identification of shift symmetry as a kind of active diffeomorphism \cite{FreidelLouapre2003,BaratinGirelliOriti2011}.
This modifies the geodesic distances between pairs of endpoints by altering the value of the metric field without altering the `position' of the endpoints themselves---we are thinking of this endpoints as the coordinate spacetime beacons discussed in the introduction.

Upon discretization, shift transformations correspond to modifications of the lengths of the edges of $\Delta$.%
\footnote{on-shell of the flatness constraint, the above modifications of the lenghts $\ell_l$ reflect displacements of the vertices of the discretization thought as locally embedded in $\mathbb R^3$.%
This is true at least for `small' displacements: in presence of boundaries and for `large' shifts, the corresponding vertex displacement might pull vertices ``out'' of the manifold \cite{ChristodoulouEtAl2012}.}
In a sense, this symmetry is imposed in Eq. \eqref{eq_PR} by `group averaging'. 
Since the group of translations is non-compact and the `gauge orbit' volume infinite, one is forced to introduce a gauge fixing---hence the prime notation.

What about the (passive) diffeomorphisms of Eq. \eqref{eq_diff}?
On a given discretization, our choice of diffeomorphism invariant variables ($h_{l^*}$ and $\ell_l$) fully takes care of them.
However, any such discretization tests only a measure zero portion of the spacetime points. 
It is possible to argue that the residual version of diffeomorphisms in this setup consists in the requirement of an invariance of the amplitude under changes in the discretization \cite{Dittrich2008,Dittrich2014solve}.

On a closed manifold, such an invariance is self-evident in the holonomy formulation (provided the discretization is fine enough to capture all the topological features of $M$). 
In the Ponzano--Regge formulation, on the other hand, it is either a consequence of its equivalence to the holonomy formulation, or---more fundamentally---of the invariance under the 2-3 and 1-4 Pachner moves---the first is nothing but the Biedenharn--Elliot identity, while the second holds in this context only formally (i.e. the equality contains an infinite prefactor).

On a manifold with boundary, however, the amplitude is a function of a certain number of boundary variables, and its  invariance under changes of the boundary discretization is a priori explicitly broken. 
In the connection polarization, flatness (and cylindrical consistency \cite{AshtekarLewandowski2004,Thiemann2004,BahrDittrichGeiller2015}) guarantee that only global discretization invariant degrees of freedom survive. 
In the metric one, flatness is explicitly broken, and the above is not the case. 
Nevertheless, upon gluing two bulk regions $M_1$ and $M_2$ across a common boundary $B_{12}$, the invariance is readily restored, since this operation requires summing over the boundary data on $B_{12}$ precisely in a way that turns the resulting amplitude equal to the amplitude for $M_1\cup_{B_{12}} M_2$.

For now, we leave a deeper study of a discretization invariant continuum limit to future work (see, however, Sec. \ref{sec_conclusions} for more comments on this). 
Instead, we focus on the identification of the quantum boundary degrees of freedom on a fixed boundary discretization and metric boundary conditions.

\section{Quantum edge modes \label{sec_edgemodes}}

After all these preliminaries, we can finally delve into the main topic of this article: the identification of the quantum edge modes of 3d gravity directly from the quantum theory.
As it will be clear soon, the edge mode theory one finds depends on the imposed boundary conditions.
For definiteness, we will perform our analysis for metric boundary conditions constructed as in Sec. \ref{sec_metricbc}.
This will allow a more direct comparison to the results summarized in the introduction.

We start from the simplest bulk topology, that of a 3-ball, $M=\mathbb B_3$ and $\pp M = \mathbb S_2$.

\subsection{Quantum Lorentz symmetry compensating fields\label{sec_Ledge}}

(This section reprises work done by the author and collaborators in \cite{ABCE1,ABCE2,ABCE3}---to which we refer for details on the following formal manipulations. With respect to that work, however, this section contains a more organic and complete discussion of the general structure of the edge mode theory.)

Putting together Eq.s \eqref{eq_ZDelta}, \eqref{eq_ampl}, and \eqref{eq_SN}, it is easy to see that these expressions can be rearranged into one involving only delta functions on the boundary dual faces%
\footnote{Since $\pp M \cong \mathbb S_2$, in this case the prime means that {\it one} redundant dual face in $\Gamma$ is omitted.}
\be
\la Z_{\mathbb B_3} | \Psi^{\mathbb S_2}_{(j,\iota)} \ra =  \Big[ \prod_{l^\ast_\pp}\int \d h_{l^\ast} \Big] \prod'_{f_\pp^*}\delta( H_{f^*} ) \Psi_{(j,\iota)}[h_{l_\pp^*}].
\label{eq_bdryeval}
\ee
The flatness condition implied by the above delta functions together with the spin-network's gauge invariance---Eq. \eqref{eq_ginv}---mean that
\be
\la Z_{\mathbb B_3} | \Psi^{\mathbb S_2}_{(j,\iota)} \ra = \Psi^{\mathbb S_2}_{(j,\iota)}[h_{l^*_\pp} = \mathbb 1] = \mathrm{tr}_\Gamma\Big[ \bigotimes_{v^\ast_\pp\in\Gamma} \iota_{v^\ast_\pp} \Big].
\ee
Notice that the two rightmost expressions above are purely boundary expressions: the bulk has been completely solved for.
This type of expression is known as a `spin-network evaluation'.

To understand what kind of edge mode theory is secretly encoded there, we observe that to obtain the last expression above we used the simple identity
\be
D^j(h=\mathbb 1)^{m'}{}_m = \delta^{m'}{}_m.
\ee
This means that the amplitude is obtained summing over a single magnetic index per vertex and depends only on the intertwiners.
Thus, we can rewrite this as
\be
\la Z_{\mathbb B_3} | \Psi^{\mathbb S_2}_{(j,\iota)} \ra = \sum_{\{m_{l^*}\}} \prod_{\{v^\ast\}} (\iota_{v^*})^{m\dots}{}_{m'\dots}
\label{eq_vmod}
\ee
(to avoid clutter we omitted the $\pp$ labels, and lowered half of the indices with the tensor $\epsilon_{m'm}$ or its spin $j$ generalization).

We claim that it is useful to interpret this expression as a (complex) statistical model, where the magnetic indices $\{m_{l^*}\}$ are the configuration variables and $(\iota_{v^*})^{m\dots}{}_{m'\dots}$ the corresponding Boltzmann weights. 
This kind of statistical models are called `vertex models', because the interaction happens around the vertices of the graph.

Somewhat equivalently, one can think of it also as a discrete (complex) path integral where the magnetic indices $\{m_{l^*}\}$ label a basis of the local Hilbert spaces (degrees of freedom) and $(\iota_{v^*})^{m\dots}{}_{m'\dots}$ are the local matrix elements of the `Hamiltonian'---of course a bit of caution has to be used, since the topology of the underlying 2d spacetime is that of the 2-sphere, $\mathbb S_2 = \pp M$.

According to the logic developed so far, the magnetic boundary degrees of freedom demand to be interpreted as the theory's edge modes---and this is how we will interpret them.
Nonetheless, in a connection picture one would more naturally expect group elements representing the local gauge transformations at the boundary to be the natural candidates for the edge modes. 
This is what happens e.g. in the derivation of the Wess--Zumino--Witten model sketched in the introduction.

To at least partially close this gap, we observe that the magnetic indices $\{m\}$ label the orientations of a quantum angular momentum vector $\vec J$ of length $\sqrt{j(j+1)}$; indeed, they constitute the most efficient orthonormal such labeling which is compatible with the uncertainty principle underlying the algebra $[J_a,J_b] = \epsilon_{ab}{}^c J_c$.
In the above descriptions, the $\{m\}$ degrees of freedom can thus be interpreted as the quantized orientations of a reference frame, which, in turn, can be classically encoded in $\SO(3)$ group elements.

The gap can now be fully closed by using an overcomplete basis of `coherent' intertwiners \cite{LivineSpeziale2007}.
A $p$-valent coherent intertwiner is labeled by $p$  $\SU(2)$ representations $V_{j_i}$, and $p$ normalized spinors $\eta_i\in\mathbb C^2\cong V_{1/2}$, $\la\eta_i|\eta_i\ra=\bar\eta^0_i\eta^0_i + \bar\eta^1_i\eta^1_i=1$. 
Each such spinor defines an $\SU(2)$ group element
\be
g_\eta = \mat{cc}{\eta^0 &  -\bar\eta^1 \\\eta^1_i & \bar\eta^0}.
\ee
Supposing all the $p$ dual edges attached to the intertwiner are outgoing, the coherent intertwiner $\iota[\eta]$ is defined, modulo a normalization factor, by
\be
\iota[j,\eta]^{n_1\cdots n_p} \sim \int_{\SU(2)} \d G \,\prod_{i=1}^p D^{j_i}(G g_{\eta_i})^{n_i}{}_{m_i = j_i}.
\label{eq_cohint}
\ee

These objects admit a beautiful geometrical interpretation in terms of (dual) quantum polygonal linkages embdeed in $\mathbb R^3$ of edge lengths given by the spins $j_i$ and edge directions $\hat v_i = \la \eta_i | \vec\sigma | \eta_i\ra$ \cite{LivineSpeziale2007, ConradyFreidel2009}.%
\footnote{There is also a 3d geometrical interpretation in terms of polyhedra of face areas given by the $j_i$. This plays a role in 3+1d loop quantum gravity. }

Without entering into the details, we just point out that the assignment of coherent intertwiners $\iota[j,\eta]$ to the dual vertices in $\Gamma=\pp\Delta^\ast$ is equivalent to the assignement of a discrete quantum metric attached to $\pp\Delta$, in perfect agreement with the picture we are developing. See \cite[Sec.IIB]{ABCE2} for details.

Plugging the coherent intertwiners of Eq. \eqref{eq_cohint} into the vertex model amplitude of Eq. \eqref{eq_vmod}, gives after some simple algebra
\be
\la Z_{\mathbb B_3} | \Psi^{\mathbb S_2}_{(j,\iota)} \ra = \Big[\prod_{v^\ast} \int_{\SU(2)} \d G_{v^\ast} \Big] e^{S_\G[G_{v^\ast}]}
\ee
where the holomorphic discrete boundary action is\footnote{The introduction of the logarithm is `artificial', and its branch cut does not introduce any ambiguity.}
\be
S_\G[G_{v^\ast}|j,\eta] = \sum_{l^\ast} 2j_l \ln [ \eta_{t(l^\ast)}  | G^{-1}_{t(l^\ast)} G_{s(l^\ast)}|\eta_{s(l^\ast)}\ra,
\label{eq_SGamma}
\ee
where $|\eta\ra \mapsto [\eta|$ is the map $\eta^A \mapsto{\eta}^B \eps_{BA}$.
Passing to the coherent basis, we have traded the sum over magnetic indices for integrals over group elements associated to the dual vertices.
Gauge transformations act simply by translating the new group-valued degrees of freedom, $G_{v^\ast} \mapsto g_{v^\ast} G_{v^\ast}$, and leave the amplitude manifestly invariant.

Thus, we found a description equivalent to the vertex model above, where not only the edge modes are compensating fields for the gauge transformations, but the fixed (metric) boundary conditions $(j,\eta)$ explicitly constitute the background structure for the edge modes theory. 
This is in complete analogy with the structure of Eq.s \eqref{eq_WZW} and \eqref{eq_edgeGR}.
For a discussion of the continuum limit of the spin-network action and its relation to the fixed-triad boundary conditions, see App. \ref{app_triadbc}. 

This description in terms of the group continuous variables plays the role of a path integral in terms of a classical action principle---albeit on a discrete spacetime---whereas the sum over magnetic indices is akin to the equivalent description in terms of matrix elements of the corresponding Hamiltonian.
In general one expects the classical theory to provide a good approximation for large quantum numbers. 
This is indeed the case: when the spins are large, $2j \gg 1$, and the magnetic indices are numerous ($m\in\{-j,-j+1,\dots,j\}$), one can use the above action principle in the stationary phase approximation to provide a good estimate of the total amplitude.%
\footnote{In this case, the coadjoint orbit corresponding upon quantization to the irreducible representation of spin $j$ is large with respect to a `Planck-sized' cell. This is what makes the classical theory a good approximation.} 
Moreover, well developed techniques allow to turn the equation of motions of $S_\G$ into geometrical statement about the (local) embedding of $\pp\Delta$ in $\mathbb R^3$ \cite{DowdallGomesHellmann2010,ABCE2, BarrettEtAl2009}, thus showing that $S_\Gamma$ is essentially an off-shell (discretized) version of the York--Gibbons--Hawking boundary term.%
\footnote{Recall, the York--Gibbons--Hawking boundary term is the integral of the boundary's extrinsic curvature. It turns the (on-shell) Einstein--Hilbert action into a differentiable functional of the induced boundary metric. Notice that the bulk part of this action, given by the Ricci scalar, vanishes on-shell of the flatness condition.}
This is a concrete version of the heuristic considerations about the correspondence between boundary states and boundary conditions put forward around Eq. \eqref{eq_heuristic}.

\subsection{Quantum shift symmetry compensating fields \label{sec_Sedge}}

The appeal of the above construction consists in having turned the amplitude of a spin-network boundary state, i.e. a spin-network evaluation, into an edge theory for Lorentz-gauge compensating fields.
Nonetheless, the original theory we started from, 3d $\SU(2)$ $BF$-theroy, featured shift symmetry as well, and no compensating field for this symmetry appears in any of the above formulations of the edge theory.

Recall, however, that we had also observed in Sec. \ref{sec_BF} that the shift symmetry is `conjugate' to the Lorentz symmetry, and it is indeed for this reason that the two do not naturally manifest at the same time.
An edge theory of shift symmetry compensating fields indeed exists and is dual to the two formulations presented so far.
It will provide us with a new quantum realization of Carlip's construction of the edge modes as `would-be normal diffeomorphisms'.

The most immediate way to find this theory is to use the Ponzano--Regge formulation of Eq. \eqref{eq_PR} on a discretization that trivializes as much as possible the role of the bulk. 
For $M=\mathbb B_3$, such a natural candidate exists and consists in choosing $\Delta$ to have single internal vertex directly connected through bulk radial edges to the boundary. 
The quantum lengths of such bulk radial edges would be the only degrees of freedom one has to sum over.
Therefore, they readily provide a quantum version of Carlip's `would-be radial diffeomorphisms' compensating fields---recall the discussion of Sec. \ref{sec_bulksym} for the relation between shift symmetry and the value of the bulk spins.

Although this derivation fully captures the substance of the shift-symmetry edge modes, it is nonetheless restricted to the case of a triangulated 3-ball.
We will now sketch a slightly different derivation, which has the advantage of being completely general and applicable to any cellular decomposition of the 3-ball and---with little adaptation---to any handlebody topology.
In particular, the focus will stay on the boundary surface, with no reference to the bulk.

To proceed, we start again from Eq. \eqref{eq_bdryeval}, but instead of using gauge invariance and solving for the delta functions to fix all $h_{l^*}$ to the identity, we rather expand the delta functions using Eq. \eqref{eq_delta} and use eq. \eqref{eq_Clebsch} to get rid of the remaining integrals.
Indeed, this is always possible because a given group element $h_{l^*}$ appears precisely in three Wigner $D^j$ matrices: one is the spin-network contribution associated to the dual edge $l^\ast$ and the other two come from the expansions of the delta functions associated to the two boundary faces it bounds,%
\footnote{The delta function omitted because of the gauge fixing can be replaced with any function of $H_{f^*}$ whose value at the identity is 1, e.g. $\chi^J(H_{f^*})/d_J$.}
$l^* = \pp f_1^*\cap \pp f_2^*$.

Then, the so-obtained amplitude reads:\footnote{All labels refer to the boundary graph $\Gamma$.} 
\begin{align}
&\la Z_{\mathbb B_3} | \Psi^{\mathbb S_2}_{(j,\iota)} \ra= \sum'_{\{J_{f^*}\}}  \prod_{v^*} W_{v^*}[J|j,\iota],\notag\\
&W_{v^*}[J|j,\iota] = \tr_{\Gamma_{v^*}}\Big( \bigotimes_{i=1}^{p_{v^*}} C^{\, j_{l^*_i}}_{J_{f^*_{1,i}} J_{f^*_{2,i}}} \otimes \iota_{v^*} \big)
\label{eq_fmod1}
\end{align}
where $\Gamma_f$ is a spin-network graph obtained by isolating a vertex $v^*\in\Gamma$ and connecting its subsequent open ends%
\footnote{At this purpose recall that $\Gamma$ is embedded in the oriented two surface $\pp M$.}
with edges labeled by spins $J_{f^*}$. See Sec. \ref{sec_graph}.

In the direct-discretization's labeling, $v^*\leftrightarrow f$, $l^*\leftrightarrow l$, and $f^*\leftrightarrow v$, this reads
\begin{align}
&\la Z_{\mathbb B_3} | \Psi^{\mathbb S_2}_{(j,\iota)} \ra= \sum'_{\{J_v\}}  \prod_f W_f[J|j,\iota],\notag \\
&W_f[J|j,\iota] = \tr_{\Gamma_{f}}\Big( \bigotimes_{i=1}^{p_{f}}   C^{j_{l_i}}_{J_{f=t(l_i)} J_{f=s(l_i)}} \otimes \iota_{f} \big).
\label{eq_fmod}
\end{align}
In these expressions, the shift symmetry compensating fields $\{J_v\}$ live at the vertices of the triangulation and geometrically represent their `radial coordinate'. 
The Boltzmann weight is a somewhat complicated quantity built out of Clebsch--Gordan coefficients contracted among themselves and with the face intertwiner (representing the shape of the face $f\in\pp\Delta$).

In Eq. \eqref{eq_fmod}, the interaction takes place around the faces (of $\Delta$), and as such defines an IRF (`Interaction Round a Face') statistical model. 
By construction, it is equivalent---or dual---to the vertex model of Eq. \eqref{eq_vmod}.

Once again, in the edge theory, the spin-network's spins and intertwiners play the role of coupling constants.

The face model formula for the spin-network evaluation is of course not new, although the physical interpretation we are proposing to the best of our knowledge is. See e.g. Turaev's `shadow calculus' \cite{Turaev1992,KirillovReshetikhin1989}, as well as \cite{Witten1989,Witten1990}.
Moreover, in \cite{BonzomDittrich2015}, a semiclassical version of the face model of Eq. \eqref{eq_fmod} was used in the study of flat-space holography, with a tentative identification of $J_v$ as Liouville-like dual fields, in analogy with Eq. \eqref{eq_edgeGR}. 

It is important to notice that the shift symmetry compensating fields identified here are those associated to `radial' displacements of the boundary. 
In this they are completely analogous to Carlip's identification of `would-be normal diffeomorphisms' with the Liouville field in AdS$_3$ \cite{Carlip2005a}.

On the other hand, diffeomorphisms tangential to the boundary surface should also play a role \cite{DonnellyFreidel2016,Geiller2017b}. 
Since the boundary spins are kept fixed by construction, one sees that that tangential diffeomorphism symmetry is explicitly broken in this setup.
We leave a discussion of this point to the closing section, Sec. \ref{sec_conclusions}.

Finally, we notice that a `first order' model where both the Lorentz and shift symmetry compensating fields appear at the same time can in principle be written by plugging  Eq. \eqref{eq_cohint} into Eq. \eqref{eq_fmod}. The ensuing expression, however, does not seem particularly enlightening---at least in that form.

\section{Solid torus and thermal field theory\label{sec_torus}}

So far we dealt with the case of a 3-ball. 
A more general case of interest, however, is that of a handlebody.
In particular, the solid torus, $S\mathbb{T}_2\cong \mathbb B_2 \times \mathbb S_1$, has a special status in that it is the background for thermal field theory, both for the 1+1 edge theory, and the 2+1 gravitational bulk theory in presence of spacelike boundaries.
For this interpretation to be consistent, the non-contractible cycle of $S\mathbb{T}_2$ has to represent the Euclidean time (inverse temperature) direction for both theories. 
This geometrical setup can be understood as a finite-space, $\Lambda=0$, analogue of the thermal AdS/CFT correspondence \cite{MaloneyWitten2007} (see also \cite[Sec. II]{ABCE1} and references therein). 

For this reasons, we will focus on the solid torus.
Generalization to arbitrary handlebodies can in principle be achieved through the same techniques.

We start again from Eq.s \eqref{eq_ZDelta}, \eqref{eq_ampl}, and \eqref{eq_SN}.
In this case, however, the delta functions and gauge invariance are not enough to fix all holonomies to the identity.
Indeed, the delta functions impose only local---not global---flatness: a non-trivial holonomy around the non-contractibe cycle of the solid torus is left.

To make the calculation more explicit, we `cut open' the solid torus $S\mathbb{T}_2$ into a solid-cylinder%
\footnote{Of course the solid-cylinder is homeomorphic to the 3-ball. We keep nevertheless this nomenclature to emphasize the role of the bottom and top disks, $\mathbb B_2 \times \{0\}$ and $\mathbb B_2 \times \{1\}$ respectively, upon gluing into a solid torus.} 
$SC_2 \cong \mathbb B_2 \times [0,1]$, in a way compatible with the triangulation $\Delta$.
In this way the cut is transverse to a set of dual edges.
Denote $R$---'ring'---the set of edges of $\pp \Delta$ along which the boundary of $S\mathbb{T}_2$ is cut, and by $R^*$ the corresponding set of dual edges of $\Gamma$.

The topology of such a cylinder is now trivial, and one can fix via flatness and gauge invariance all the holonomies---except those associated to dual edge in $R^*$---to the identity, as in the previous sections. 
Local flatness forces the remaining holonomies---those associated to $l^*\in R^*$---to be all equal to some $g\in\SU(2)$, which is simply the holonomy around the non-contractible cycle of $S\mathbb{T}_2$.
Integration over all possible locally flat bulk holonomies implicit in Eq.s \eqref{eq_ZDelta} leads to 
\begin{align}
\la Z_{S\mathbb T_2} |& \Psi^{\mathbb T_2}_{(j,\iota)} \ra= \notag\\
&= \int \d g\, \Psi^{\mathbb T_2}_{(j,\iota)}[h_{l^*\notin R^*}=\mathbb 1, h_{l^*\in R^*} = g]\notag\\
& =\tr_\Gamma \Big[  \bigotimes_{v^\ast} \iota_{v^\ast} \otimes \mathbb H_{R^*} \Big],\label{eq_bdryevalT2}
\end{align}
where $\mathbb H_{R^*}$ is an operator acting on the dual edges in $R^*$:
\be
\mathbb H_{R^*} = \int \d g \bigotimes_{l^\ast\in R^*}  D^{j_{l^*}}(h_{l^\ast\in R^*}=g).
\ee

Equation \eqref{eq_bdryevalT2} emphasizes that the local theory is the same as above, modulo the insertion of an extra operator, $\mathbb H_{R^*}$.

The operator $\mathbb H_{R^*}$ is nothing but a so-called Haar intertwiner, which decomposes simply as
\be
\mathbb H_{R^*} = \sum_I | \iota_I \ra \la \iota_I|,
\ee
with $I$ labeling an orthonormal basis of $r$-valent intertwiners, $r=\#R^*$.

The location of the ring $R$ is completely irrelevant as a consequence of local flatness and gauge invariance. 
As a consequence, $\mathbb H_{R^*}$ is a `topological operator' for the edge theory.

\section{Example:\\ spin 1/2 quadrangulation of the torus and integrable models \label{sec_onehalf}}

Consider now the case of a cellular decomposition $\Delta$ of $S\mathbb T_2$ such that $\pp \Delta$ is a quadrangulation of $\mathbb T_2$, and fix the boundary conditions to be those imposed by a spin-network with all spins $j_l = 1/2$ \cite{ABCE1}. 

The quadrangulation has $T$ `time-like' edges `parallel' to the non-contractible cycle, and $L$ horizontal `space-like' edges `parallel' to the contractible one.
Space-like edges are dual to `vertical' dual edges, and time-like edges to `horizontal' dual edges.
A `twist' of $N_\gamma$ units can be inserted before identifying back the space-like edges belonging to the ring $R$. The twisting angle will be 
\be
\gamma = 2\pi\f{N_\gamma}{L}.
\ee
The ratios $T/L$ and $N_\gamma/L$ constitute the modulus of the torus.

The space of 4-valent intertwiners between spin 1/2 representations is 2 dimensional.
In fact, choosing an arbitrary recoupling channel ($s$, $t$, or $u$) for its decomposition onto two 3-valent intertwiners, the recoupling spin can take only the values 0 or 1, e.g. $|\iota \ra = \alpha |s=0\ra + \beta |s=1\ra$, $\alpha,\beta\in\mathbb C$.
A more convenient basis is provided by picking two different spin-zero recoupling channels $|\iota \ra = \lambda |s=0\ra + \rho |u=0\ra$, or for brevity of notation
\be
|\iota[\alpha,\rho] \ra = \alpha |s\ra + \beta |u\ra.
\label{eq_0int}
\ee
In components,\footnote{In components, $|t\ra$ reads $\eps^{m_1 m_4}\eps_{m_2 m_3}$.}
\be
\iota[\alpha,\beta]^{m_1 m_4}_{m_2 m_3} = \alpha \delta^{m_1}_{m_2} \delta^{m_4}_{m_3} + \beta \delta^{m_1}_{m_3}\delta^{m_4}_{m_2},
\ee
where $(m_1, m_2)$ are indices on the horizontal dual edges, while $(m_3,m_4)$ on the vertical dual edges.

In the conventions of \cite{Faddeev}, 
\be
\iota[\alpha,\beta] = \f{i\beta}{2}L(\lambda),
\quad
\lambda =  \f{\alpha}{-i\beta}+\f{i}{2},
\label{eq_Lax}
\ee
where $L(\lambda)$ is the Lax operator for the isotropic Heisenberg spin-chain, or XXX spin-chain
The parameter $\lambda$  is called the `spectral parameter'.

The XXX spin-chain is the isotropic version of the XXZ spin-chain, a protitypical example of a quantum integrable system.
The Hamiltonian of the periodic XXZ spin-chain acts on the Hilbert space of $L$ spins 1/2, $\cH_L=V_{1/2}^{\otimes L}$, and is given by
\be
H_\text{XXZ} = -\f14 \sum_{n=1}^L \Big(\sigma^1_n \sigma^1_{n+1} + \sigma^2_n \sigma^2_{n+1} + \Delta \sigma^3_n \sigma^3_{n+1}\Big),
\ee
where $n+L \equiv n$ labels the sites of the chain, and $a=1,2,3$ the three space directions. 
The XXX Hamiltonian is obtained in the isotropic limit $\Delta =1$.
Integrable higher spin generalizations also exists, but the integrability condition makes their Hamiltonian is more complicated \cite{Faddeev} (see below).

As a 1+1 quantum integrable system,  the XXZ spin-chain is equivalent to a 6-vertex model with Boltzmann weights $(a,b,c)$ such that
\be
\Delta = \f{a^2+b^2 - c^2}{2ab}.
\ee

The spectral parameter above maps onto the following Boltzmann weights for the `isotropic' (i.e. $\Delta=1$) version of the 6-vertex model
\footnote{Or, equivalently, to $a = \alpha +\beta$, $b=\alpha$, $c=\beta$.}
\be
a = \alpha, 
\quad 
b = \alpha+\beta,
\quad
c = \beta.
\ee
The equivalence of the edge theory of Eq. \eqref{eq_vmod} for the 4-valent spin 1/2 case to the isotropic 6-vertex model can also be found by simple inspection, mapping the $m=\pm1/2$ degrees of freedom onto arrow directions \cite{ABC1}.

In particular, it is immediate to see that---modulo an unimportant overall normalization factor---each space-like slice of the edge theory provides a copy of the isotropic spinchain (or 6-vertex model) transfer matrix%
\footnote{We keep following the notation of \cite{Faddeev}.}
\be
F(\lambda) = \tr_\h\Big[ \bigotimes_{n=1}^L L_n(\lambda) \Big] : \cH_L \to \cH_L.
\label{eq_F}
\ee
Here, $\tr_\h$ means that only the magnetic indices corresponding to the horizontal dual edges have been contracted.
$F(\lambda)$ is a polynomial of order $L$ in $\lambda$.

The origin of integrability is to be found in the Yang--Baxter equation satisfied by the Lax operators
\begin{align}
&R_{\h_1,\h_2}(\lambda-\lambda_2) L_{n,\h_1}(\lambda_1) L_{n,\h_2}(\lambda_2) \notag\\
&=L_{n,\h_2}(\lambda_2) L_{n,\h_1}(\lambda_1)R_{\h_1,\h_2}(\lambda-\lambda_2)
\label{eq_YB}
\end{align}
where the labels $(\h_1,\h_2,n)$ explain that multiplication among the Lax operators takes place along the vertical dual edges at a given site $n$ of the chain, while the $R$-matrix,
\begin{align}
& R(\mu):V_{1/2}\to V_{1/2}, \notag\\
& R(\mu)^{m n'}_{n m'} = \mu  \delta^{m}_{n} \delta^{n'}_{m'} + i \delta^{m}_{m'}\delta^{n'}_{n},
\end{align}
contracts along the horizontal dual edges.

A direct consequence of this equation is that $[F(\lambda), F(\lambda')]=0$ for any value of $\lambda$ and $\lambda'$.
In turn, this means that the coefficients in $F(\lambda)$ of $\lambda^p$, $p=0,\dots,L$, commute among them. 
Among these coefficients\footnote{Actually, the following quantities appear as combinations of these coefficients.} one finds  the XXX Hamiltonian $H_\text{XXX}$, the 1-site translation operator $U$, and the chain's total spin
\be
\vec S = -\f{i}{2}\sum_{n=1}^L \vec \sigma_n.
\ee
Consequently, this allows to identify as many conserved charges as degrees of freedom (integrability).

We mentioned the total spin charge explicitly, because, when the ring $R$ of Sec. \ref{sec_torus} coincides with one space-like slice of $\pp\Delta$, the insertion of the Haar operator is equivalent to that of a projector onto the vanishing total-spin sector of the chain, i.e.
\be
\mathbb H_{R^*} = \mathbb P_{S=0}.
\ee

Putting all this ingredients together, we can finally express the edge theory partition function on $\Delta$ in the transfer matrix language as%
\footnote{We neglect an overall, unimportant, normalization factor.}
\be
\la Z_{S\mathbb T_2} | \Psi^{\mathbb T_2}_{(j=\f12,\iota[\alpha,\beta])} \ra  = \tr_{\cH_L}\Big[ F(\lambda)^T  U^{N_\gamma} \mathbb P_{S=0} \Big].
\ee

Since in the vertex model representation of the edge theory we have found a well-known integrable model, it is worth investigating its face model dual.

The latter turns out to be the (somewhat degenerate) isotropic limit of another well-known (of course integrable) IRF model of the SOS type.
This acronym stands for `Solid On Solid', and is meant to allude to the growth of a surface. 
Curiously the gravitational interpretation fits this physical picture: here the surface in question is the spacelike boundary of a portion of spacetime and its growth happens in the radial direction. 

Notice that the difference between two neighboring (radial) heights $J_v$ is necessarily $1/2$,%
\footnote{This is often renormalized to 1 via the obvious change of variables $J\mapsto 2J$.}
since $\delta_{j=1/2,J_v^1, J_v^2}$ in Eq. \eqref{eq_Clebsch} would vanish otherwise.

The correspondence between the 6-vertex and RSOS model is a well known one \cite{PasquierEtiology,KirillovReshetikhin1989}, and goes beyond what we presented here to incorporate the more general 8-vertex model (see e.g. \cite{WestonEtAl}). 
However, it is interesting to notice how our framework casts this correspondence in terms of a Fourier duality between two edge theories associated to the two conjugate gauge symmetries of 3d quantum gravity.

In working out the partition function in the solid torus case, the only subtlety one has to deal with is the translation of the Haar operator $\mathbb H_{R^*}$.
Recall that $\mathbb H_{R^*}$ is the operator that tells the boundary theory which cycle of the torus is contractible in the bulk---the boundary delta functions that we `Fourier transformed' to get to Eq. \eqref{eq_fmod} impose local flatness only, and have no global information.
For this we refer to App. \ref{app_HaarRSOS}.

Extensions beyond the spin 1/2 case that preserve integrability exist.
They are known as `descendent' models (see e.g. \cite{PinkBook}), and essentially consist in restricting to homogeneous and isotropic 4-valent spin-network states characterized by a spin $j$ and intertwiners of the form of Eq. \eqref{eq_0int} \cite{Witten1989}.
In fact, the use of general intertwiners%
\footnote{Here we are using the terminology common in loop quantum gravity. In the integrable model literature, by `intertwiner' one often means an $R$-matrix, while here we generally call `interwiner' something more akin to an $S$-matrix (scattering matrix) of two spin $j$ quasi-particles scattering among themselves, possibly exchanging fundamental spin-chain excitations of spin 1/2. }
would break integrability since it would not correspond to a Lax operator satisfying a Yang--Baxter, as in Eq.s \eqref{eq_Lax} and \eqref{eq_YB}.

Staying with the spin 1/2 case, correspondences with non-isotropic models are also possible, provided the gravitational theory is modified by the addition of a cosmological constant. 
Before delving into this subject we present in the next section a graphical notation that will simplify our task---and possibly clarify what we have accomplished so far.

\section{Graphical notation\label{sec_graph}}

To a certain cellular decomposition of the boundary $\pp\Delta$ (in black) we associate its Poincar\'e dual $\Gamma=\pp\Delta^*$ (in red)
\begin{center}
\includegraphics[width=.4\textwidth]{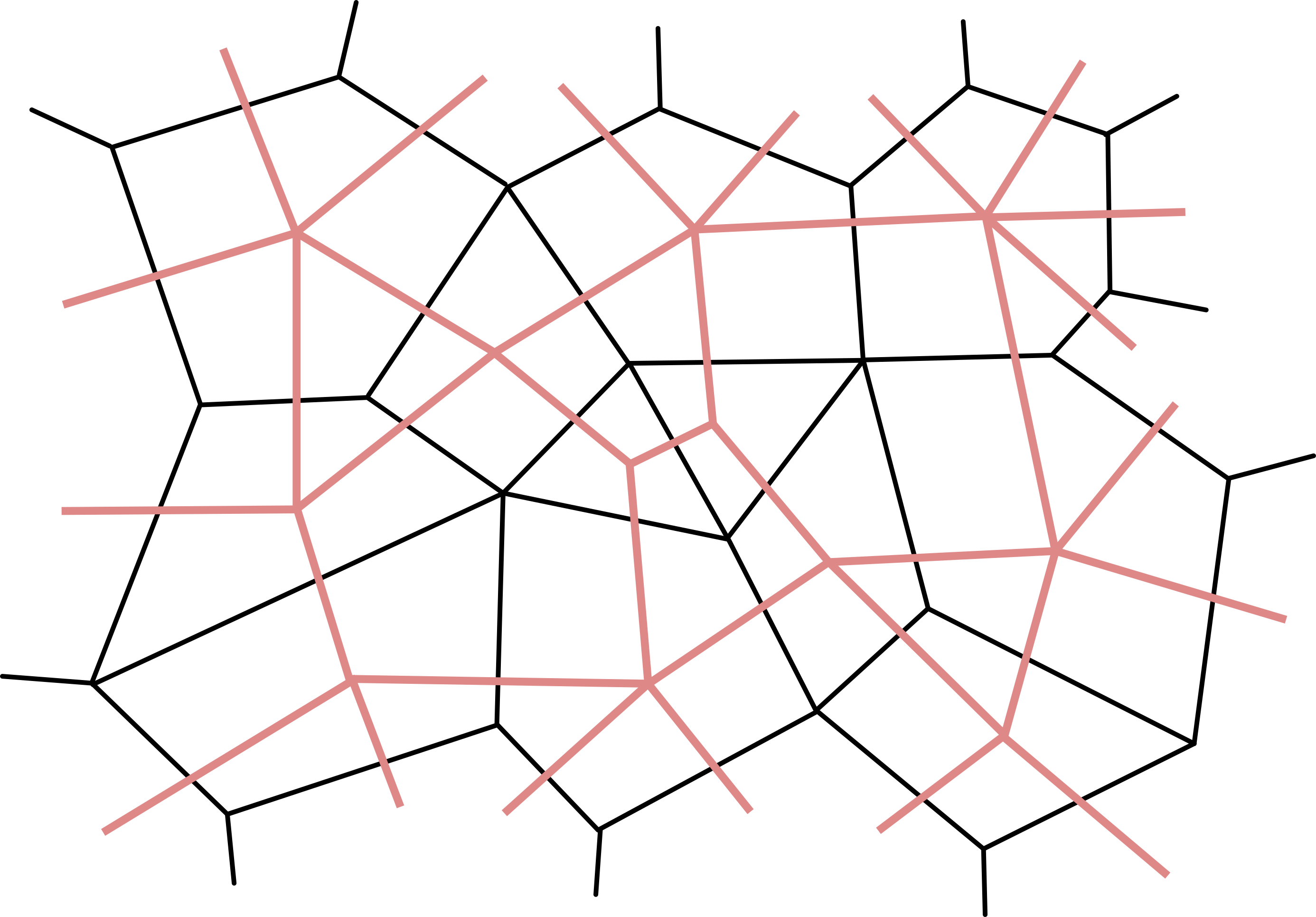}
\end{center}

To define the spin-network function $\Psi_{(j,\iota)}[h_{l^*}]$---which imposes metric boundary conditions to the gravitational amplitude---we first associate to dual edges (red lines) $l^*\in\Gamma$ labeled by a spin $j_{l^*}$ the Wigner matrix (composition is from left to write),
\be
D^j(h)^m{}_{m'} =\;\raisebox{-.3em}{ \includegraphics[scale=.3]{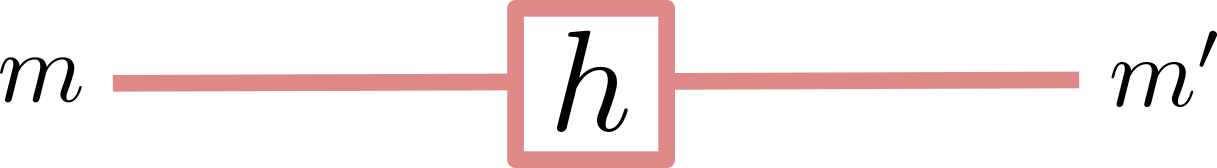} },
\ee
and to dual vertices (intersections of red lines) the intertwiner $\iota_{v^*}$ (all dual edges are outgoing)
\be
\iota^{m_1,\dots,m_p} =\;\raisebox{-1.5em}{ \includegraphics[scale=.3]{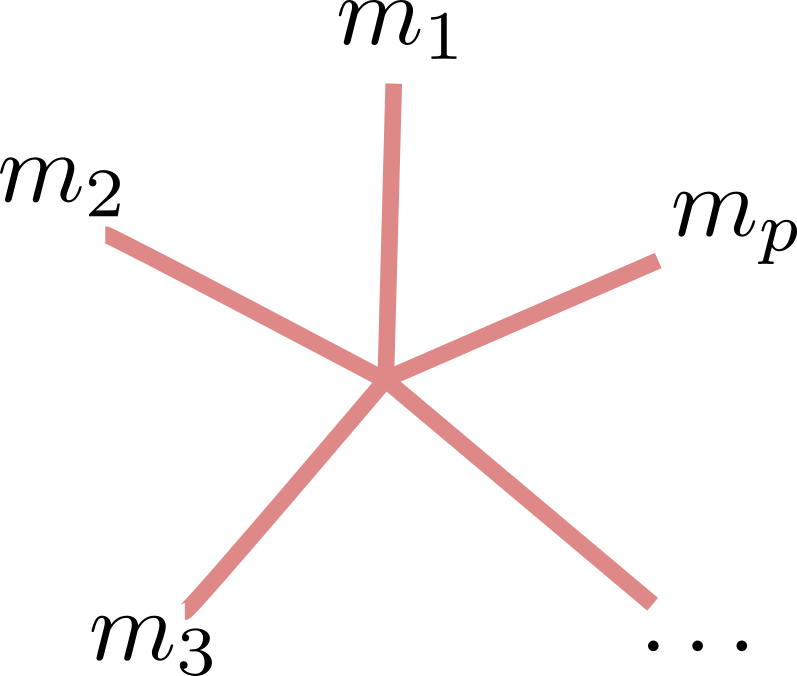}} \;.
\ee
The spin-network function is finally obtained by contracting%
\footnote{Contraction between two upper indices is done with the spin $j$ generalization of the $\SU(2)$ invariant tensor $\epsilon_{mm'}$, i.e. $(-1)^m\epsilon_{m,-m'}$. See \cite{BarrettNaishGuzman2009} for details.} 
all the magentic indices $m$ according to the combinatorics imposed by $\Gamma$ (Eq. \eqref{eq_SN}):
\be
\Psi^\Gamma_{(j,\iota)} =\; \raisebox{-4.8em}{ \includegraphics[scale=.3]{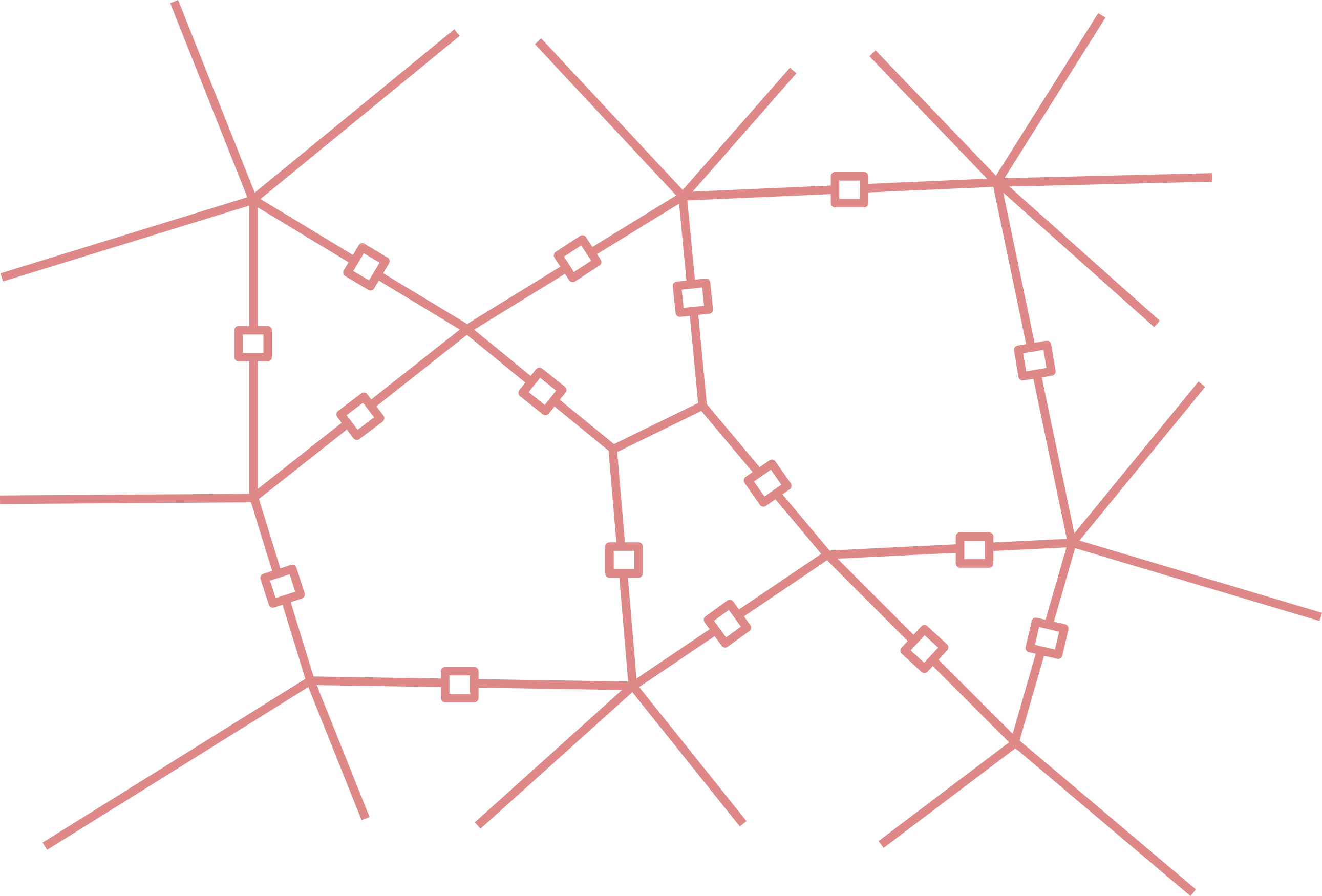}}
\ee

We then represent the delta function on $\SU(2)$ by a dashed line (Eq. \eqref{eq_delta})
\be
\delta(h) = \;\raisebox{-1.2em}{ \includegraphics[scale=.3]{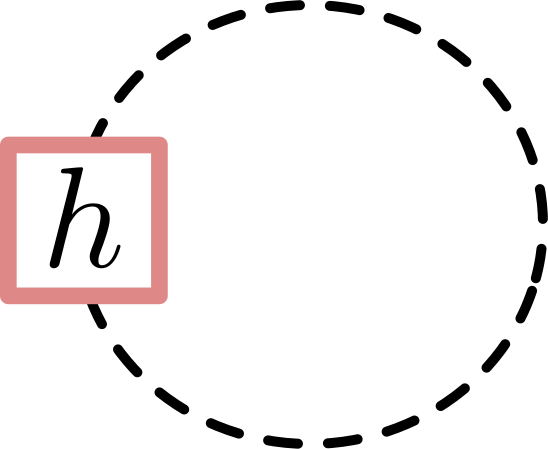}} = \sum_{J} (-1)^{2J}d_J \raisebox{-1.2em}{ \includegraphics[scale=.3]{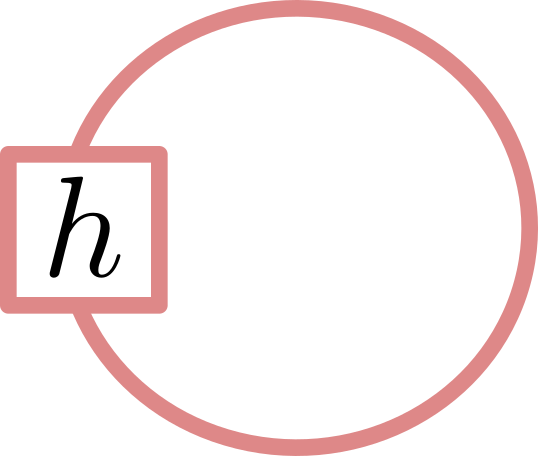}}.
\label{eq_deltafgraph}
\ee

Denoting integration over a common variable by a box crossing multiple dual edges, i.e.
\be
\raisebox{-1.6em}{ \includegraphics[scale=.3]{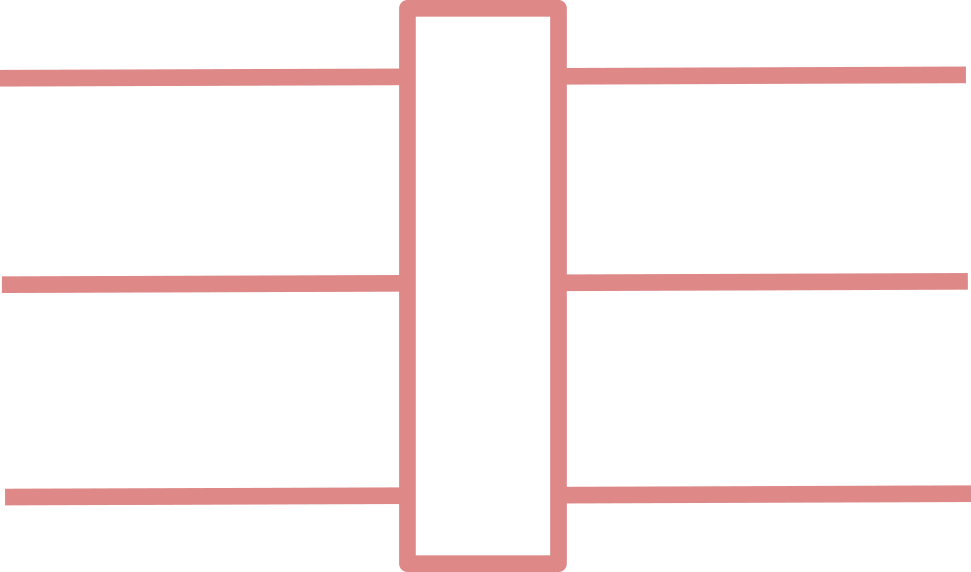}}\;=\;\int \d g \;\raisebox{-1.6em}{ \includegraphics[scale=.3]{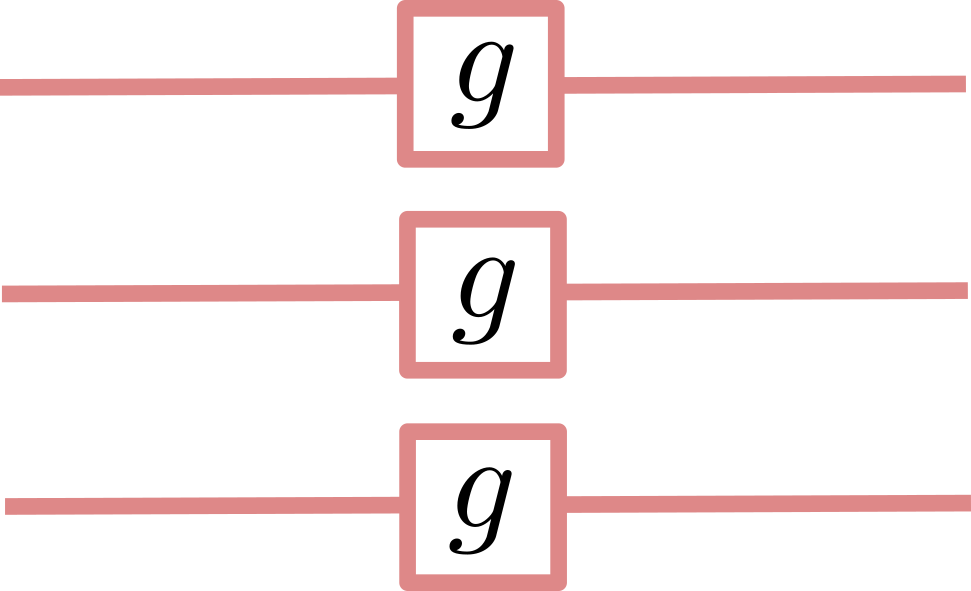}}\;,
\ee
we write Eq. \eqref{eq_Clebsch} as 
\be
\raisebox{-1.6em}{ \includegraphics[scale=.3]{integral.png}}\; = \;\raisebox{-1.2em}{ \includegraphics[scale=.3]{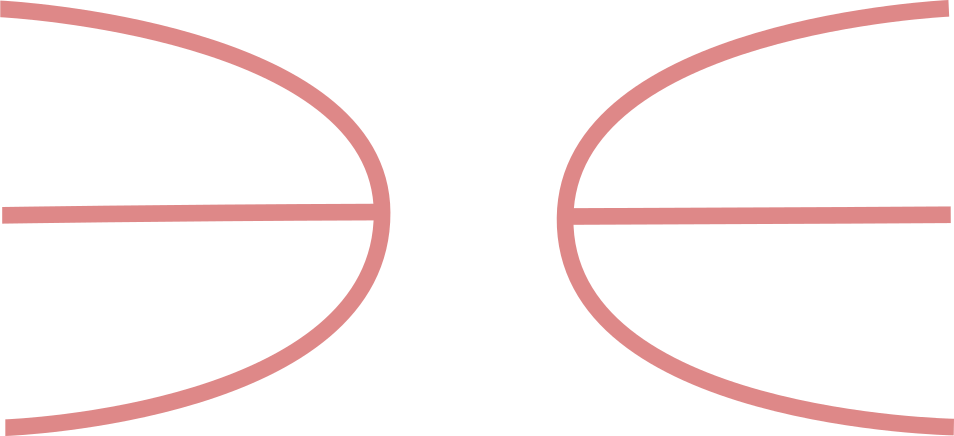}}\;.
\ee

Thus, if $M=\mathbb B^3$ and $\pp\Delta$ is a cellular decomposition of $\pp M = \mathbb S_2$, the spin-network evaluation of Eq. \eqref{eq_bdryeval} can be represented as
\begin{align}
&\Big[\int \d h_{l^*} \Big] \prod_{f^*} \delta(H_{f^*}) \Psi^\Gamma_{(j,\iota)}[h_{l^*}] = \notag\\
& =\; \raisebox{-4.8em}{ \includegraphics[scale=.3]{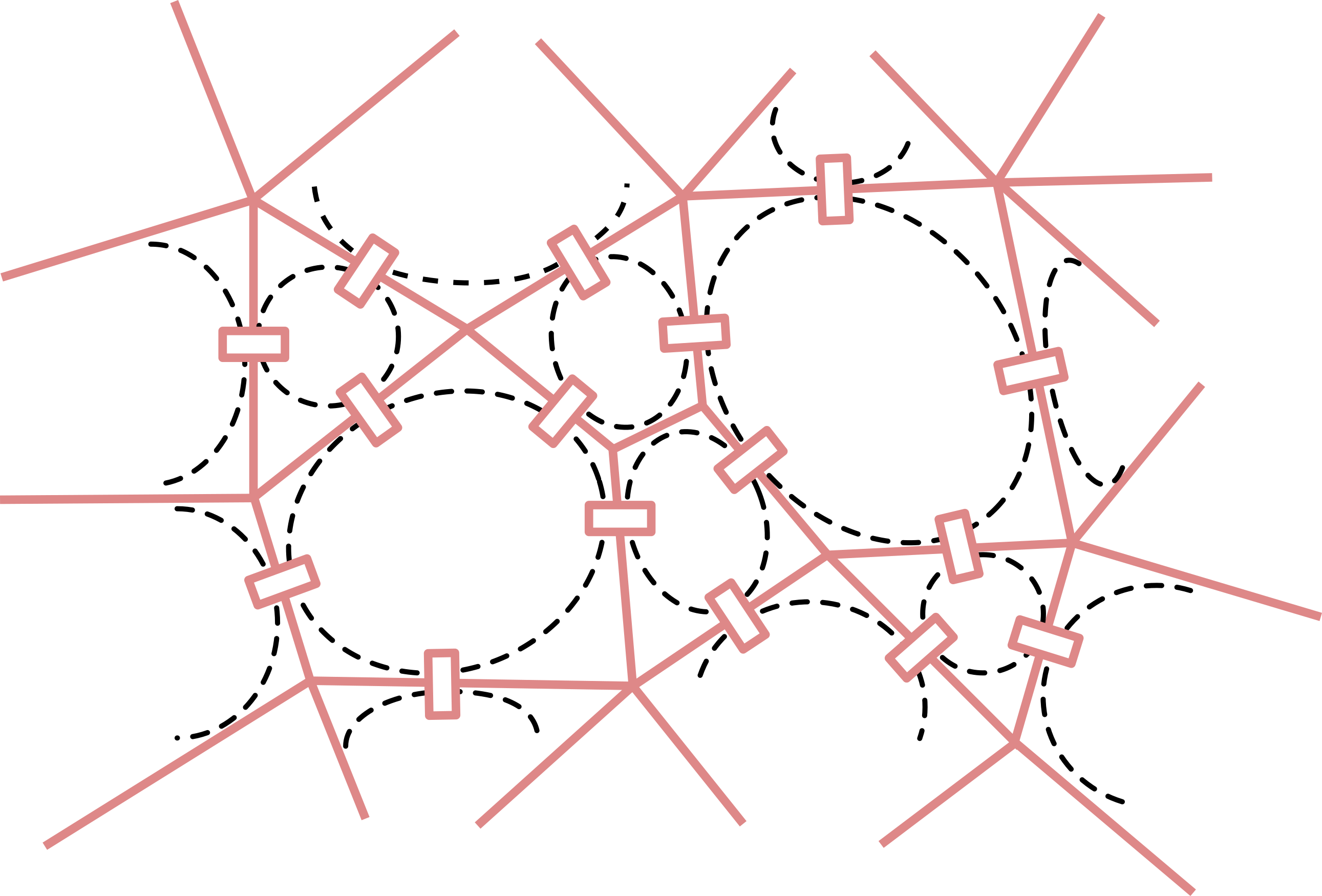}}
\label{eq_deltaf2}
\end{align}
For a general topology, this equation imposes on $\Psi_{(j,\iota)}$ the local flatness condition for the boundary surface $\pp M$, with no reference to the bulk topology.

Using the graphical calculus described above, Eq. \eqref{eq_deltaf2} can be turned into the following graphical expression
\begin{align}
\raisebox{-4.8em}{ \includegraphics[scale=.3]{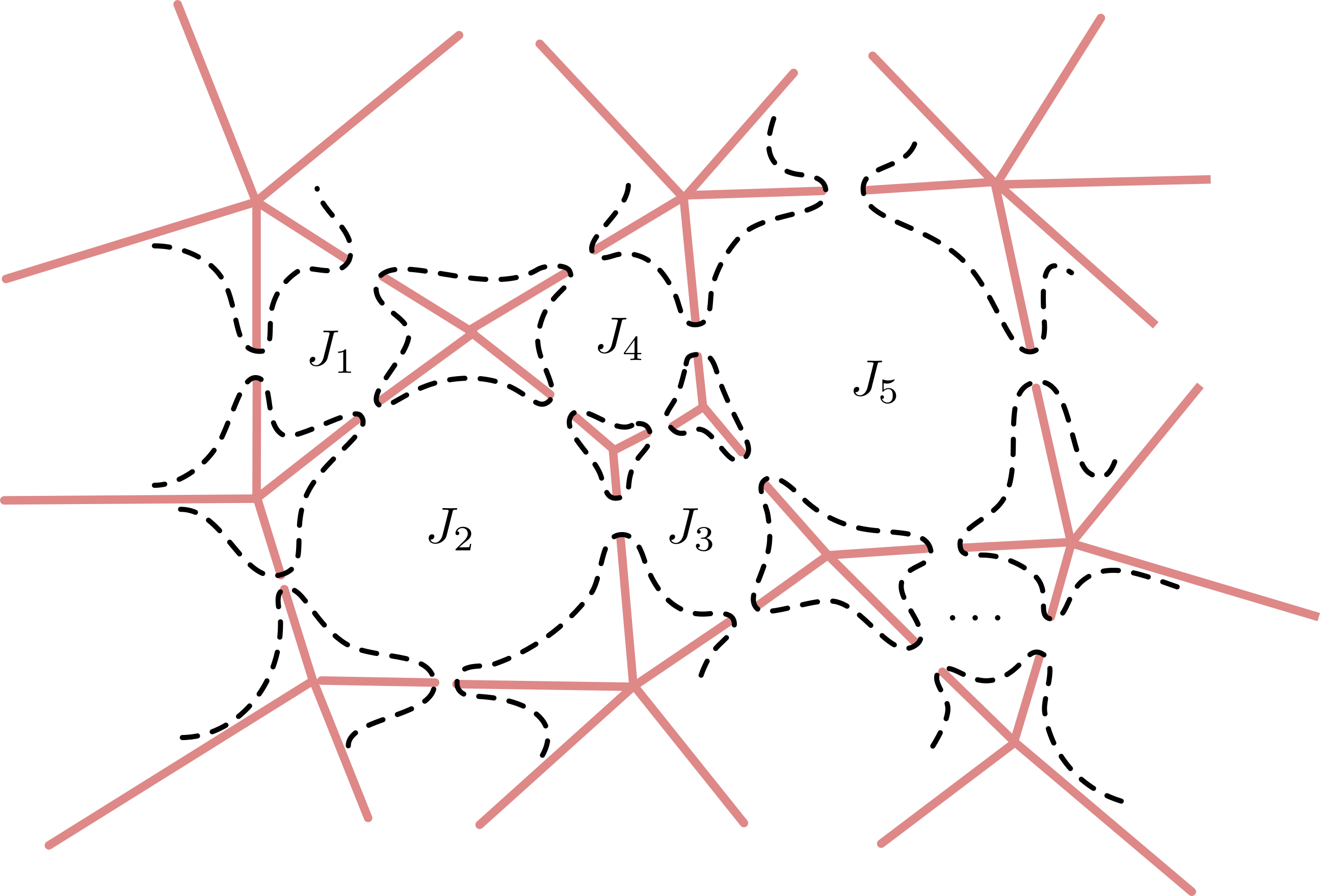}}
\label{eq_deltafrec}
\end{align}
Here we kept the dashed-line notation with a spin $J_{f^*}$ at the center of each dual face to underline which spins are summed over, as well as to remind the reader that the coefficients appearing in Eq. \eqref{eq_deltafgraph} are left understood.

Around each dual vertex there is a local graph $\Gamma_{v^*}$.
As contractions of intertwiners and Clebsch--Gordan coefficients, they represent the weights of Eq. \eqref{eq_fmod1},
\be
W_{v^\ast}[J|j,\iota] = \;\raisebox{-.8em}{ \includegraphics[scale=.3]{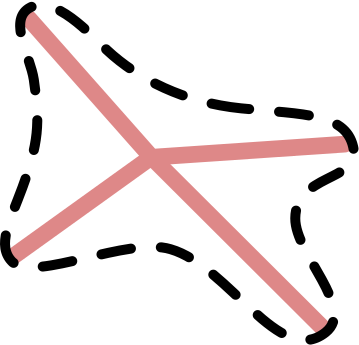}}\;.
\ee
E.g. the 3-valent case evaluates to a $6j$-symbol (we refer to \cite{BarrettNaishGuzman2009} for a careful treatment of the normalizations of these expressions)
\be
W_{v^*}[J|j,\iota] \;
= \;\raisebox{-.4em}{ \includegraphics[scale=.3]{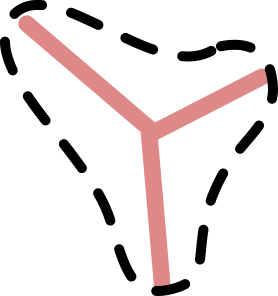}}\;
\sim\; \left\{
\begin{array}{ccc}
j_1 & j_2 & j_3 \\
J_4 & J_5 & J_6 
\end{array}
\right\}\,.
\ee

In this way, the right hand side of Eq. \eqref{eq_deltafrec} represents graphically the IRF model of Eq. \eqref{eq_fmod1}.
A maybe more transparent notation is given in terms of $\Delta$:
\be
\raisebox{-4.8em}{ \includegraphics[scale=.3]{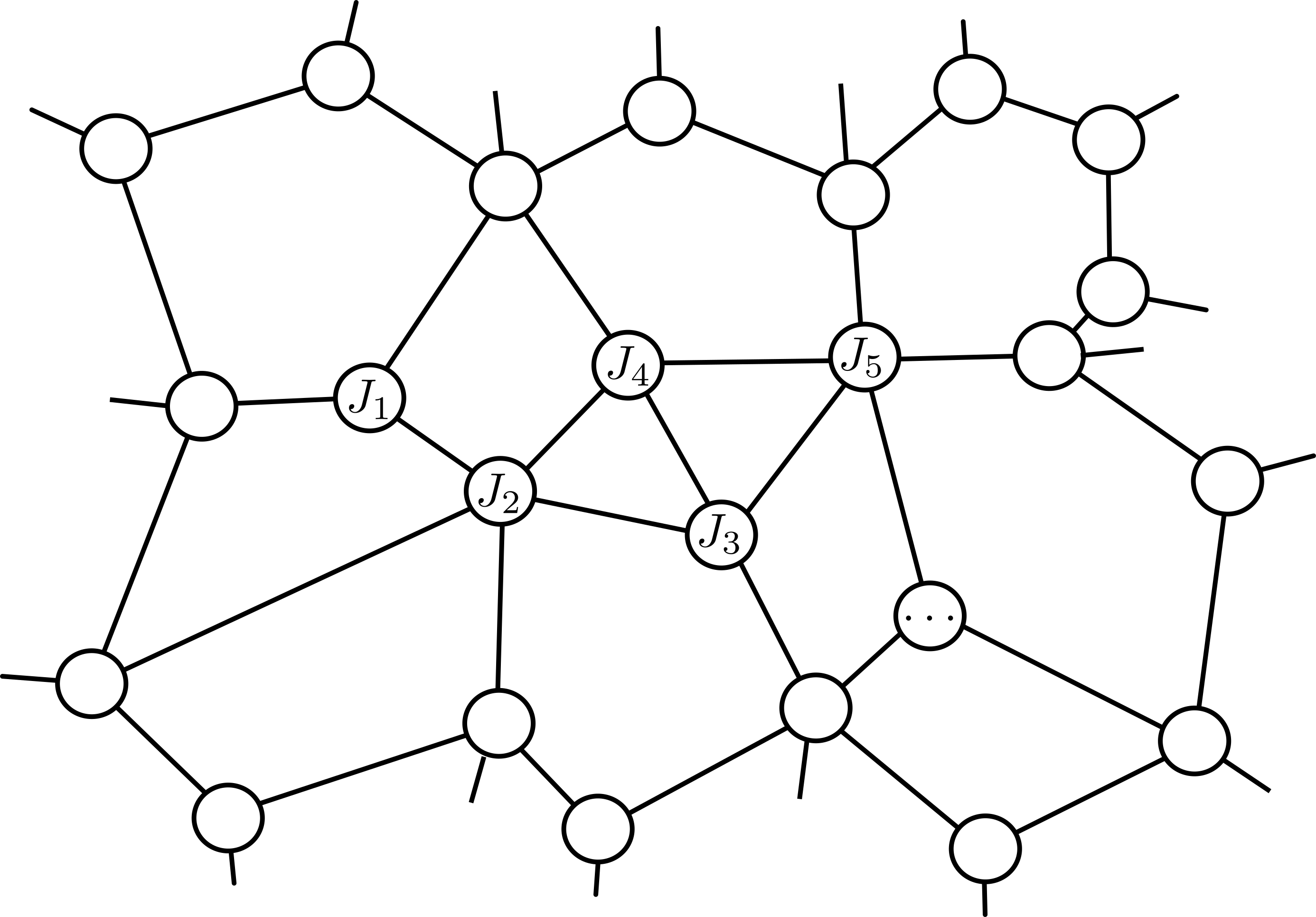}}
\ee
where the edges $l$ (black lines) carry a spin $j_l$ representing their lengths, and the faces carry the intertwiners $\iota_f$ representing their shapes (recall that on the boundary $l\leftrightarrow l^*$, and $f\leftrightarrow v^*$) as well as the IRF weights of Eq. \eqref{eq_fmod}, e.g.
\be
W_{f}[J|j,\iota] = \;\raisebox{-2.4em}{ \includegraphics[scale=.3]{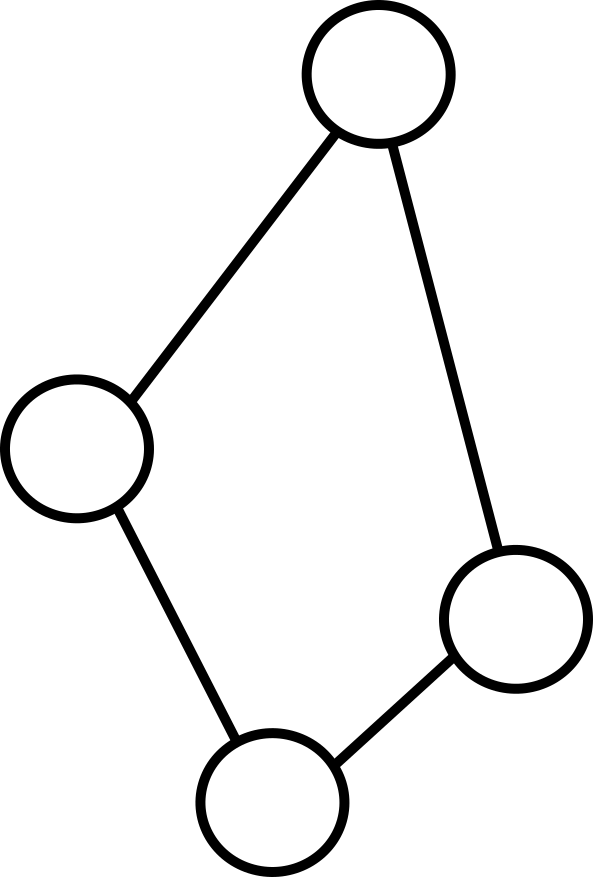}}\;.
\label{eq_IRFgraph}
\ee
or
\be
W_{f}[J|j,\iota] \;
= \;\raisebox{-1.em}{ \includegraphics[scale=.3]{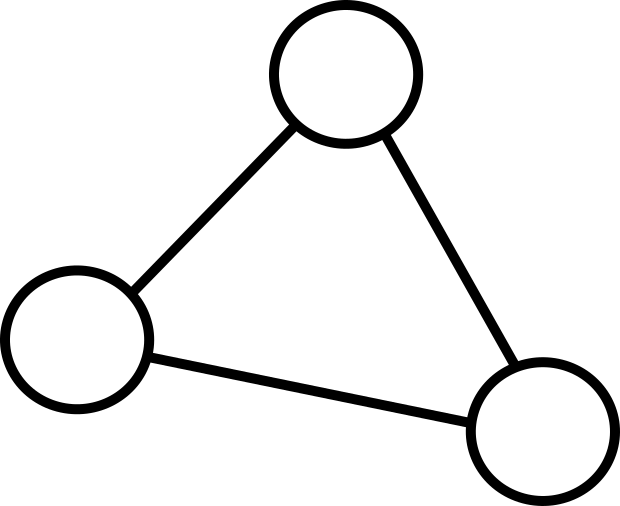}}\;
\sim\; \left\{
\begin{array}{ccc}
j_1 & j_2 & j_3 \\
J_4 & J_5 & J_6 
\end{array}
\right\}\,.
\ee 
The spins in the circles are the variables one needs to sum over.
Geometrically they represent the distance of a vertex of $\pp\Delta$ from some (fiducial) point in the bulk.
They constitute the quantum shift symmetry compensating field.

\section{Cosmological constant and the Turaev--Viro model\label{sec_cc}}

In presence of a cosmological constant $\Lambda$ the first-order action is
\be
S_\omega = \f1\ellpl \int \delta_{ab} e^a\wedge F^b[\omega] - \f{\Lambda}{3!} \epsilon_{abc}e^a\wedge e^b \wedge e^c,
\ee
and the equations of motions are
\be
F^a \EOMeq \f\Lambda2 \epsilon^a{}_{bc}e^b\wedge e^c,
\quad
\D_\omega e^a \EOMeq 0,
\ee
that is constant curvature (rather than flatness) and torsion-freeness, respectively.

It is immediate to see that while Lorentz symmetry is untouched, transformations of the form of Eq. \eqref{eq_flatshift} are not symmetries anymore.

At the level of the cellular complex, this happens because in this case the vertex translations that shift symmetry induces must take place in an homogeneously curved space, rather than in flat space.
In fact, if $\Lambda \neq 0$, the nature of shift symmetry is modified:
\be
\delta_\lambda e = \D_\omega \lambda,
\quad
\delta_\lambda \omega =  - \Lambda\ad_\lambda e,
\ee
where both $\omega$ and $e$ are here considered as $\su(2)$ valued. 

A better way to deal with this is to notice that the total internal symmetry group is now deformed from $\ISU(2)$ into
\be
G_\Lambda \cong
\begin{cases}
\SL(2,\mathbb C)  & \text{if } \Lambda <0  \\
\SO(4)  & \text{if } \Lambda >0
\end{cases}
\ee
Accordingly, one can set $\omega = \omega^a J_a$ and $e = e^a P_a$, where the translation generators $P_a$ are now deformed to boost (or `Euclidean boost') generators,
\be
[P_a,P_b] = \Lambda \eps^{c}{}_{ab} J_c.
\ee
This allows to phrase the theory in a form more similar to Eq. \eqref{eq_Somega} \cite{Witten1988}.
Of course, in this setting one also sets $X = X^aJ_a$ and $\lambda = \lambda^a P_a$.

The groups $G_\Lambda$ can also be assigned a (quasi-)Poisson--Lie structure, which puts into evidence the two conjugate parts of the symmetry group in analogy to Eq. \eqref{eq_ISO}. These are of course rotations and (Euclidean) boosts.
However, since the boosts do not constitute a group, the treatment is more involved and we restrain from detailing it here---see e.g. \cite{Bonzom2014,DupuisGirelliLivine2014,Riello2017}.

One important aspect is that, from the perspective of the symmetries of the theory, $e$ and $\omega$ are now on much more similar footing, and a successful discretization must take this into account.
The resulting (canonical) quantization, which is also more subtle, leads to a (lattice) Hopf-algebra gauge theory \cite{MeusburgerWise2016,Meusburger2017} which `deforms' the lattice gauge theory construction which implicitly underlaid our discussion of spin-network states.

If $\Lambda >0$, the resulting Hopf-algebra gauge theory is essentially a Kitaev model \cite{Kitaev2003} for the $U_q(\SU(2))$ Turaev--Viro code \cite{TuraevViro1992,Kirillov2011,BalsamKirillov2012,Meusburger2017}, with\footnote{Recall that in our definition $\ellpl = 8\pi G_\text{N} \hbar$. Here $\ell_\text{c} = 8\pi/\sqrt{\Lambda}$ can be interpreted as the scale of the cosmological horizon.} 
\be
q=e^{\f{2\pi i}{k+2}},
\quad
k = \f{8\pi}{\ellpl \sqrt{\Lambda}} \in \mathbb N_+.
\ee 

From a gravitational perspective, the Turaev--Viro state sum model is a deformation of the Ponzano--Regge model of Eq. \eqref{eq_PR}.
In particular, the asymptotics of the $q$-deformed $6j$-symbol reproduces the Einstein--Hilbert--Regge action of a {\it positively curved} tetrahedron in presence of a cosmological constant \cite{MizoguchiTada1992,TaylorWoodward2005}.
This confirms the above intuition and assigns to the $j$'s the interpretation of geodesic lengths in a curved spacetime.%
\footnote{If $\Lambda<0$, and $q\in(0,1)$ is real, the asymptotics of the $6j$ symbol still reproduces the expected Einstein--Hilbert--Regge action for a negatively curved tetrahedron. 
However, the resulting Ponzano--Regge-like model is plagued by divergences.
We shall not consider this case any further, even if the following considerations can be adapted to this case too.}

Hence, the whole construction of the previous sections can be directly generalized by replacing spin-network evaluations with $q$-deformed ones.
In particular the graphical formulas of Eq.s \eqref{eq_deltafrec}  and \eqref{eq_IRFgraph} preserve their validity once all relevant symbols are appropriately $q$-deformed.

Therefore, it should not come as a surprise that that the discussion of Sec. \ref{sec_onehalf} on the spin 1/2 case also admits a $q$-deformed generalization.
The resulting 6-vertex and RSOS models (as well as their dualities) are discussed in \cite{PasquierEtiology} (see also \cite[Sec. 5.2]{Witten1989}). 
We will not delve into the details of these models, and will simply emphasize that there the cosmological constant shows up in the form of a non-trivial anisotropy parameter $\Delta$. Explicitly:
\be
\Delta = \f12( q + q^{-1} ) = \cos\left( \f{2\pi}{k+2} \right).
\ee

\section{Canonical picture\label{sec_canonical}}

So far we have worked in an (Euclidean) covariant picture, which allowed us to deal with all boundaries in the same way, regardless on whether they are `space-like' or `time-like'.
It is however instructive to look at the canonical picture too. 

In this section, we will have to attribute a different interpretation to some of the notation introduced above.
We will emphasize when this happens.

The geometrical set up is now that of a manifold of the form $M = \Sigma\times[-\varepsilon,\varepsilon]$, i.e. a collar neighborhood of a `space-like' surface $\Sigma$.
For clarity, but committing an abuse of language, we will refer to $\Sigma$ as the `Cauchy surface'.
The infinitesimal `time-like' boundary surface, will be denoted  $B = C \times [-\varepsilon,\varepsilon]$, where $C = \pp \Sigma$ stands for `corner'. 
For definiteness, we shall restrict to the case where $M$ is a `solid cylinder', and hence $\Sigma\cong\mathbb B_2$, $C \cong \mathbb S_1$.

We discretize $\Sigma$ via a cellular decomposition $\Delta_2$---the subscript `2' emphasizes the 2-dimensional nature of the cellular complex, in contrast to the notation used in the rest of the paper.
Let $\Delta_2^*$ be the Poincar\'e dual of $\Delta$, and denote it by $\Gamma = \Delta_2^*$---in this, section $\Gamma$ strictly refers to the discretization of  the Cauchy surface $\Sigma$.

\subsection{Closed Cauchy surface: $\pp \Sigma = \emptyset$}

Let us start from the case of a closed Cauchy surface $\Sigma$, $\pp \Sigma=\emptyset$.

If $\Lambda=0$, in order to quantize the theory \`a la Schroedinger we can then proceed similarly to Sec. \ref{sec_quant}: we first smear $A=\underleftarrow \omega$ on dual edges $l^*\in\Gamma$ to obtain a finite set of parallel transport variables $h_{l^*}$, and then we build the Hilbert space $\cH'_\Gamma$ of $L^2$ functions of these variables:\footnote{Although the notation is the same as in Sec. \eqref{sec_quant}, there $\cH_\Gamma$ did not have strictly speaking the interpretation of a Hilbert space.}

On $\cH'_\Gamma$, two sets of constraints act. These are the discrete version of the Gauss (Lorentz) and flatness (shift) constraints of Eq. \eqref{eq_constr}.
The first can be imposed by group averaging and reduces $\cH'_\Gamma$ to its gauge-invariant counterpart $\cH_\Gamma$
\be
\Psi[h_{l^*}] \in \mathcal H_\Gamma = L^2\Big( \SU(2)^{\times L^*} // \SU(2)^{\times V^*}\Big).
\ee
$\cH_\Gamma$ is the Hilbert space of a $\SU(2)$ lattice gauge theory.
A basis is provided by the spin-network states $\Psi^\Gamma_{(j,\iota)}$.

The flatness constraint is in turn imposed by projecting on those states whose support is restricted to configurations such that (see Eq. \eqref{eq_Hfast})%
\footnote{Strictly speaking this procedure is not a `projection', due to the measure zero character of the flat configurations. See \cite{AshtekarLewandowski2004,Thiemann2004} for details.}
\be
H_{f^*} = \mathbb 1.
\ee

In the gravitational parlance, the imposition of the flatness constraint reduces $\cH_\Gamma$ to the `physical' Hilbert space $\cP_\Gamma$.

Let us now compare with the language used in the Kitaev model literature.
There, the Gauss constraint is imposed by the action of the $A$ operator, which is interpreted as annihiliating the { electric} flux out of a face $f\in\Delta_2$.
Similarly, the flatness constraint is imposed by the $B$ operator, which is in turn interpreted as annihilating the { magnetic} flux through a dual face $f^*\in\Delta_2^*$.
Finally, the analogue of $\cP_\Gamma$ corresponds to the ground state (vacuum) sector of the model.

In the rest of this section, I will stick to the gauge theoretic electric-magnetic language, rather than the gravity oriented triad-connection one.

The continuum limit of $\cP_\Gamma$ can be obtained either \`a la loop quantum gravity via an inductive limit construction \cite{Baez1994,AshtekarLewandowski2004,Thiemann2004}, or \`a la spin-net via the introduction of equivalence classes of graphs \cite{LevinWen2005}---in contrast to finite groups or quantum groups with a finite Rep category, for Lie groups the spin-net construction is much less natural \cite{DittrichGeiller2015,BahrDittrichGeiller2015,ABC1}.

If $\Lambda>0$, the classical phase space analogue of $\cH_\Gamma$ is a deformation of the symplectic quotient $ \mathrm T^*\SU(2)^{\times L^*} // \SU(2)^{\times V^*}$ of the form $\SO(4)^{\times L^*} // \SU(2)^{\times V^*}$ (see \cite{Riello2017} for details). 
Its quantization and reduction to the flat sector, however, is most easily expressed in a spin-net picture for the finite Rep category $\mathcal C=\mathrm{Rep}(U_q\SU(2))$, $q$ root of unity---e.g. \cite{LevinWen2005,DittrichGeiller2017,Dittrich2017}

\subsection{Corners: $\pp \Sigma = C \neq \emptyset$}

So far, the role of the corner $C=\pp \Sigma$ has been neglected. 
The first question one needs to answer regards the following discrete ambiguity: how does $C$ cut through the edges of $\Delta_2$ and the dual edges of $\Delta_2^*$?
In other words, which one among $\pp\Delta_2$ and  $\pp\Delta_2^*$ is a discretization of $C$? 
Since $\Delta_2$ and $\Delta_2^*$ naturally carry either electric or magnetic excitations respectively, the above question is indeed one of physics. (See also \cite{CasiniHuertaRosabal2014}). 

For Kitaev's models, magnetic boundary conditions have been studied in detail, e.g. \cite{BombinDelgado2008,BeigiShorWhalen2011}.
Consistently with the rest of the paper, we will here rather focus on electric type boundary conditions.
The following is a representation of the Cauchy surface $\Sigma\cong\Delta_2$ (in dark blue), of its bounding corner $\pp\Sigma=C\cong\pp\Delta_2$ (in black), as well as of the dual discretization $\Gamma=\Delta_2^*$ (in light blue):
\begin{center}
\vspace{.5em}\includegraphics[scale=.3]{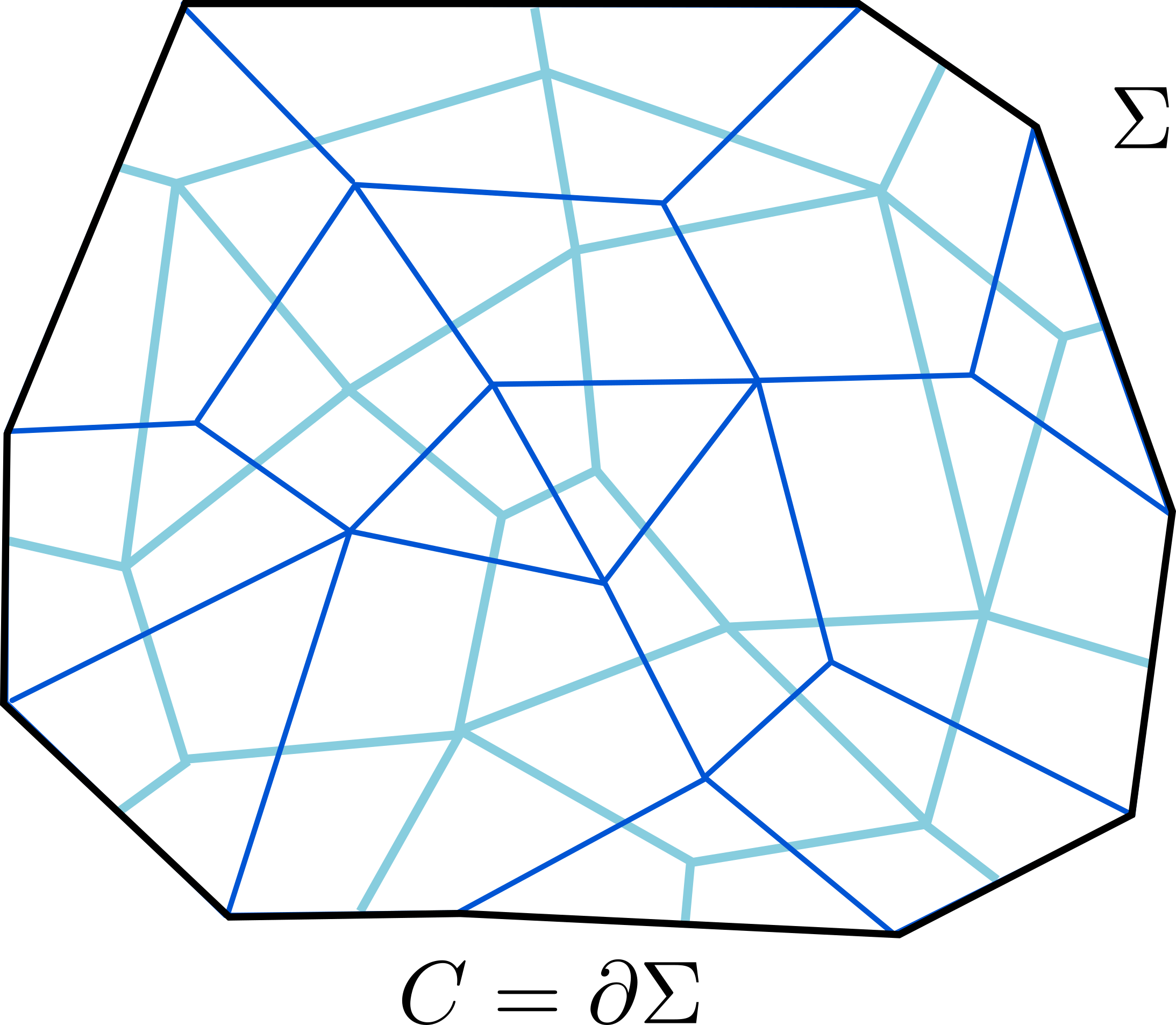}
\end{center}

The electric boundary conditions we want to impose consist of fixed spins along the (black) boundary edges.

To identify the edge modes, we first observe that an `open' dual edge ends at each edge of $C$.
Gauge invariance cannot be imposed at those open ends without trivialiazing the information they carry and thus hindering the possibility of gluing a region back to its complement.
This fact implies that in presence of corners $C\neq\emptyset$, boundary magnetic indices $\{m_{l^*_C}\}$ have to be added to the count of degrees of freedom.
Their Hilbert space is
\be
\cH^\text{gauge}_{\pp \Sigma} = \bigotimes_{l\in \pp\Delta_2}V_{j_l}.
\ee
A natural expectation is that these are the (Lorentz) gauge symmetry compensating fields.

To confirm this expectation, one can match them with the construction of the Sec. \ref{sec_Ledge}, which gave a covariant treatment of the fields at the `time-like' boundary $B$.
At this purpose, we represent here a portion of the `time-like' boundary $B$  of $M$  (in black) and its dual (in red):
\begin{center}
\vspace{.5em}\includegraphics[scale=.3]{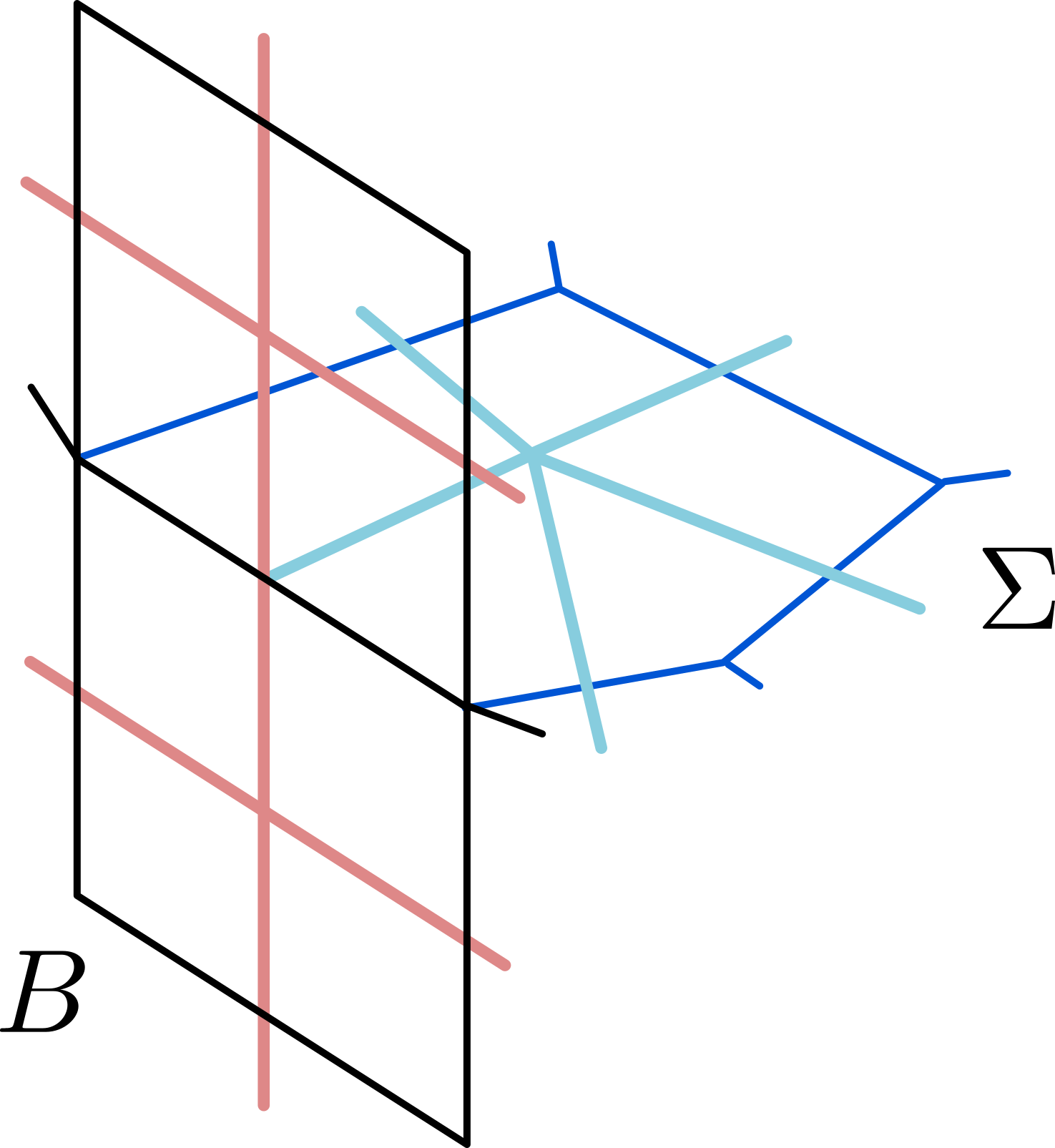}
\end{center}

From this picture it is clear that the `canonical edge modes' live precisely at the $B$-boundary dual edges (in red) exactly as it was found in Sec. \ref{sec_Ledge}.
Their dynamic is dictated by the details of the (electric) boundary conditions at $B$, i.e. by the spins and intertwiners associated to the black (or red) edges lying in  $B$. 
Gravitationally, this is akin to a coupling of the edge modes to the induced boundary metric on $B$, which our boundary conditions demand to be fixed.

Notice also that a row of square $B$-boundary faces---as in
\begin{center}
\vspace{.5em}\includegraphics[scale=.3]{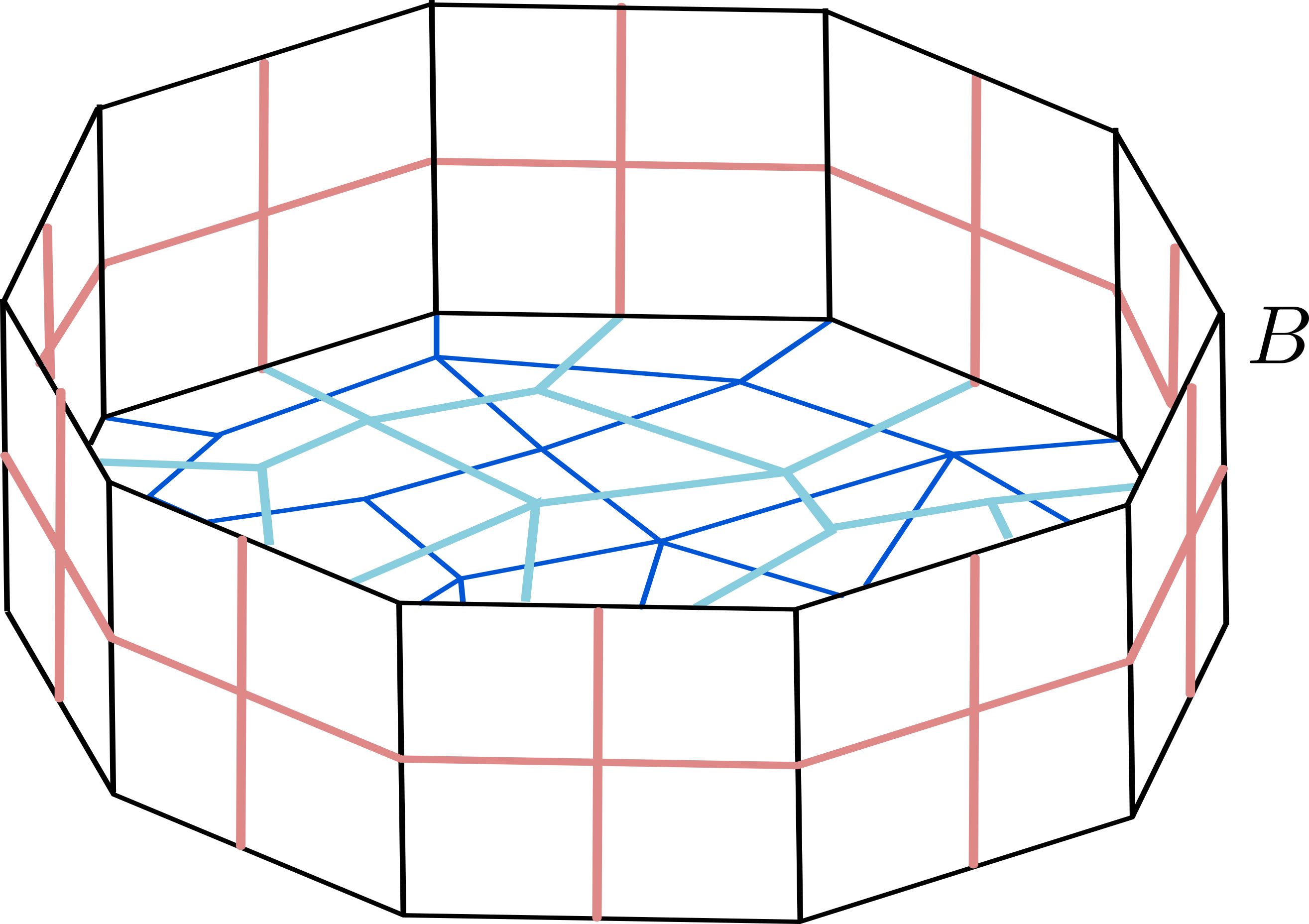}
\end{center}
---provides through its dual spin-network $\Gamma_B$ a transfer matrix 
\be
F_B(j,\iota) : \cH^\text{gauge}_{\pp \Sigma} \to \cH^\text{gauge}_{\pp \Sigma},
\ee
representing a 1-step time-like evolution of the gauge edge modes.
For the spin 1/2 boundary conditions of Sec. \ref{sec_onehalf}, this is precisely the XXX spin-chain transfer matrix%
\footnote{In this case, $\cH^\text{gauge}_{\pp\Sigma}$ is precisely the $\cH_L$ of Eq. \eqref{eq_F}.}
$F(\lambda)$ of Eq. \eqref{eq_F}.

Tracing back the manipulations of Sec. \ref{sec_Sedge}, it is easy to see that according to the dual view where the boundary degrees of freedom are the lengths (spins $J_v$) of the edges in $\Delta_2$ reaching the corner $C\subset B$ at the vertices of the discretization of $B$.
Summing over these boundary degrees of freedom implements the flatness of the connection around dual faces in the discretization of $B$. 
This statement is not associated to a single `time slice' $\Sigma$, but rather to properties of its time evolution.

\subsection{Interfaces and gluings}

To conclude our analysis, let us comment on the situation where $C$ is an interface at which two regions get glued to each other.

Consider a Cauchy surface $\Sigma$, which can be closed $\pp\Sigma=0$, and a line $C\cong \mathbb S_1$ dividing it into two regions.
In this setting, rather than a boundary, $C$ is an interface between two subsystems, $\Sigma = \Sigma_A \cup_C \Sigma_B$.
To each subregion $\Sigma_{A,B}$ we can apply the construction above. 

Beside the gauge symmetry at the end of the dual edges piercing $C$, also the flatness (zero-magnetic flux) constraint is broken for those dual faces of $\Delta_2^*$ cut by $C$ into two dual `half-faces'.
This is because, as a consequence of the uncertainty principle, the fact of fixing the electric flux through the (black) boundary edges of $\pp\Delta_2$, automatically prevents us to have control over the magnetic fluxes through the dual half-faces of $\Delta_2^*$ bounded by these same edges.
Here, the one in question is the {\it canonical} flatness associated to faces lying on $\Sigma$, rather than on $B$ as above.

At $C$, both gauge and shift invariance are restored when sewing back $\Sigma_A$ and $\Sigma_B$ into $\Sigma$.
In particular, gauge invariance is restored by summing over the edge modes $\{m_{l^*_C}\}$, while shift invariance is restored by summing over all possible boundary conditions $j_{l_C}$.

Had we chosen magnetic, rather than electric, boundary conditions, we would have found a dual setup: spins $j_{l^*_C}$ would have been interpreted as the boundary degrees of freedom compensating for a broken shift symmetry for the faces lying on $\Sigma$ and cut by $C = \pp \Delta_2^*$,%
\footnote{From the condensed matter perspective this is a new effective symmetry of the vacuum sector. It is at the origin of the topological nature of the gapped vacuum phase. }
while the gauge group elements%
\footnote{Recall the discussion of Sec. \ref{sec_Ledge} where we showed that the sum over magnetic indices can be replaced by integrals over group elements. This is the most appropriate choice here, because $C=\pp\Delta_2^*$ intersects the boundary at dual edges and dual vertices of the dual discretization of $B$.}
$G_{v^*}$ would have been interpreted as the fixing of the boundary conditions. 

This interface picture is particularly pertinent when computing entanglement entropies between subregions of $\Sigma$. 
In this context, the role of the edge modes and it s relation to the boundary conditions has been already largely emphasized e.g. in  \cite{Donnelly2008, CasiniHuertaRosabal2014, SoniTrivedi2016, DonnellyWall2015, ABCE2}.
Where comparison is meaningful, these treatments agree with ours in the edge mode identification.

\section{Summary and Outlook\label{sec_conclusions}}

In this paper we have analyzed the nature of the quantum edge modes for three dimensional quantum gravity as a Ponzano--Regge--Turaev--Viro topological field theory with metric boundary conditions.
From a guage-theoretical perspective this corresponds to the study of the edge modes in a topological sector of a non-Abelian gauge theory with electric boundary conditions

Paying attention to the smearing of the triad (electric field) and connection (magnetic potential) along dual cellular decompositions, and to the conjugate nature of Lorentz (gauge) and shift symmetries (an effective symmetry), we have unveiled a pair of dual formulations of the edge mode theory.

The first formulation is in terms of a vertex-type statistical model whose configuration variables are some magnetic indices labeling a basis in an irreducible representation of the gauge group. 
We showed how to translate these configuration variables into honest group elements representing the (Lorentz) gauge frame at the boundary, in a Wess--Zumino--Witten-like fashion.
As we observed in the last section, the magnetic-index edge modes match independent constructions performed in the study of interfaces in relation to the computation of the entanglement entropy for gauge theories.

The second formulation is in terms of face-type statistical model whose configuration variables are irreducible representations (spins) attached to the vertices of the discretization. 
These edge modes are the compensating fields for the broken shift symmetry.
 We argued that the gravitational interpretation is in term of the quantum (discrete) analogue of Carlip's `would-be normal diffeomorphisms', which he showed to reproduce the Liouville field at the boundary of AdS$_3$.

For the simplest example of metric (electric) boundary conditions, these two models gives rise to the celebrated duality between a six vertex (or XXZ spin-chain) and RSOS face models.
 
Furthermore, we discussed in some simple examples how the topology of the bulk reflects on the edge theory. 
We also pointed out, in an appendix, how our construction seems strictly related to other proposed spinorial edge theories.

Finally, we see two main---but intertwined---directions in which our investigation can be further pushed.
On the one hand, it seems necessary to understand the symmetries of our edge theories.
The role of such symmetries has been emphasized on quite general grounds in a number of discussions performed in the continuum, e.g. \cite{Elitzur1989, Balachandran1992, Balachandran1996, DonnellyFreidel2016, Geiller2017a}.
On the other hand, to fully match these continuum treatment, it is of paramount importance to better understand how to take a continuum limit in our setup.
This topic leads us to one last detour.

In all our discussion, the boundary spins are kept fixed by construction, since they encode the sought metric (electric) boundary condition.
A consequence of this fact is that shift symmetry in the tangential direction is explicitly broken in this setup.
Following the arguments reviewed in Sec. \ref{sec_bulksym}, it is possible to argue that restoration of this symmetry should be related to an invariance under changes in the boundary discretization (diffeomorphism symmetry).
Although amplitudes with this properties exist, e.g. \cite{KarowskiSchrader1993,Arcioni2001}, they are essentially spin-variable rewritings of pure connection boundary conditions.
Obtaining a similar result for metric boundary conditions is more subtle, and we expect it to involve some tuning to a second order phases transition of the boundary theory (see also the conclusion section of \cite{ABCE3}).

Of course, the mapping onto statistical models performed above can be of great advantage in addressing the previous two questions, at least in the simple cases related to thoroughly studied integrable models---e.g. it is known that the effective continuum description of an XXZ spin chain is done in terms of Wess--Zumino--Witten model, whose symmetries are well-understood; nonetheless, a more careful and detailed analysis is needed to confirm any (too) naive expectation---but also suggests that  the nature of the continuum limit might be influenced by the chosen graph connectivity. We leave all further investigations of these matters to future work.

\acknowledgements
The content of this paper was put together during a series of travels that brought me to the U. of Nottingham, the Heriot--Watt U. of Edinburgh, and the ENS Lyon. It is therefore a pleasure to thank John Barrett, Bernd Schroers, and Etera Livine for their hospitality and various fruitful conversations, Robert Weston for his kind patience, and Des Johnston for his generous invitation.
It is also a pleasure to thank my colleagues at Perimeter Institute, in particular Bianca Dittrich,  William Donnelly and Wolfgang Wieland, for discussing with them prompted me to clarify some central points of this work.
This work is supported by Perimeter Institute for Theoretical Physics. Research at Perimeter Institute is supported by the Government of Canada through Industry Canada and by the Province of Ontario through the Ministry of Research and Innovation.

\appendix
\section{Fixed-triad boundary conditions\label{app_triadbc}}

In the $BF$ formulation of 3d gravity, fixed-triad boundary conditions require the following boundary term
\begin{subequations}
\begin{align}
S_e[\omega,e] &= \f1\ellpl\int e\wedge F + \f1\ellpl\oint E\wedge A, \label{eq_Se}\\
\dd S_e[\omega,e] &\EOMeq \f1\ellpl\oint \dd E \wedge A. 
\end{align}
\end{subequations}
This boundary term equals the integral of the trace of the extrinsic curvature as in Eq. \eqref{eq_GRact}, provided the appropriate gauge is chosen (i.e. such that $\pp_\mu n^a = 0$, where $n^a = e^a_\mu n^\mu$, and $^\mu$ is the tangent space unit vector orthogonal to the boundary) \cite{
 Obukhov1987, ArcioniBlau2002, AshtekarLewandowski2004}.

Clearly, the action $S_e$ fails to be Lorentz or shift invariant, and moreover its gauge variations fail to be proportional to a constraint,
\begin{subequations}
\begin{align}
\delta_X S_e & = \f1\ellpl\oint X \d E,\\
\delta_\lambda S_e & = \f1\ellpl\oint \lambda (\d A - F).
\end{align}
\end{subequations}
This makes even a formal quantization in the triad polarization quite awkward.

Let us focus on the Lorentz gauge symmetry. 
In our discrete covariant treatment of Sec. \ref{sec_edgemodes}, it was never broken.
This was because the boundary conditions were imposed by coupling to a spin-network functional, which was gauge invariant by construction.
From this construction we were also able to read off the boundary action of Eq. \eqref{eq_SGamma}, i.e.
\begin{align}
S_\G[G_{v^\ast}|j,\eta] & =\sum_{l^\ast} 2j_l \ln [ \eta_{t(l^\ast)}  | G^{-1}_{t(l^\ast)} G_{s(l^\ast)}|\eta_{s(l^\ast)}\ra \notag\\
& = \sum_{l^\ast} 2j_l \ln [ \eta_{t(l^\ast)}  | h_{l^*}|\eta_{s(l^\ast)}\ra,
\end{align}
where in the last equation we have emphasized that the $G_{v^*}$ just encode a (globally) flat connection,
\be
h_{l^*} =  G^{-1}_{t(l^\ast)} G_{s(l^\ast)}.
\ee

We want to take a formal continuum limit of this expression, one in which the holonomies are small,
\be
h_{l^*} \approx \mathbb 1 + A^a_\mu \d (l^*)^\mu \tau_a,
\ee
$\tau^a = -\frac{i}{2}\sigma^a$.
For this we need to recall that the holonomies $h_{l^*}$ are computed along dual edges of the triangulation, {\it transverse} to the direct edges of which the spins $j$  and spinors $\eta$ are the lenghts and directions.
Hence, labeling $\mu=1,2$ directions along the boundary locally adapted to $l^\mu$ and $(l^*)^\mu$, we see, somewhat sloppily, that $S_\Gamma$ is rather the discretization of an action of the form
\begin{align}
S_\G \sim & \oint 2j_1 [  \eta_1| \pp_1 +  A_1 |\eta_1\ra  +  2j_2 \la  \eta_2| \pp_2 +  A_2 |\eta_2\ra\notag\\
\sim & \oint 2 [ j\eta_t| \d \wedge |\eta_s\ra + \f1\ellpl E^a \wedge  A_a,
\end{align}
where, because of the dualization in the cellular decomposition, (this equation has {\it no} sum over repeated indices)
\be
\ellpl j_\mu\la \eta_\mu | \sigma^a |\eta_\mu\ra  = \epsilon_{\mu\nu}E^a_\nu \d x^\nu,
\ee
(in this formula, we used the matching condition $|\eta_s\ra = |\eta_t]$, see \cite{ABCE2}).

In the continuum, a more sensible version of this action can be obtained by breaking the symmetry between source and target spinors through the introduction of the following boundary action:
\be
\oint {\underline\lambda}^\dagger \wedge (\d+ A) \eta
\label{eq_bdryW}
\ee
where $\eta \in \mathbb C^2$ and ${\underline{\lambda}}\in \Omega^1(\pp M, \mathbb C^2)$, i.e.
\be
{\underline{\lambda}} = \lambda_\mu \underleftarrow{\d x}^\mu, \quad \lambda_\mu \in \mathbb C^2.
\ee
This action is complex, therefore one has to take e.g. minus its imaginary part.
Now, the equation of motion for the connection, spurring from both the bulk and boundary contributions to the action, couples the bulk to the boundary degrees of freedom by requiring
\be
E^a_\mu \EOMeq \ellpl \Re \la \lambda_\mu |\sigma^a |\eta\ra.
\ee
Keeping the above combination of spinors fixed, through this equation of motion the boundary action above plays precisely the role of  the $\ellpl^{-1}E\wedge A$ term discussed at the beginning of this section, while preserving Lorentz-gauge invariance.

In any case, we see that loosely speaking the spin-network action manages to be Lorentz-gauge invariant by modifying the boundary term of Eq. \eqref{eq_Se} through the introduction of a spinor fields which have to identified with the `square root' of $E$ (thus, in a sense, the quartic root of the metric). 
Once the spinors are introduced a natural Lorentz-covariant boundary differential can be used.

The action above was firstly introduced in a Plebanski formulation of four-dimensional gravity in \cite{Wieland2016} (see also \cite{Wieland2017a,Wieland2017b}). 
In three dimensions, the same author put forward another proposal  for a spinorial edge-mode theory \cite{Wieland2017c}.
There, a single spinorial field appears in the action, accompanied by a `background' one-form $q^a$ intrinsic to the boundary $\pp M$ (essentially a fiducial value for $E$).

The boundary action of \cite{Wieland2017c} can be obtained from that of Eq. \eqref{eq_bdryW} by demanding
\be
|\lambda_\mu\ra\equiv q_\mu^a\sigma_a |\eta\ra.
\ee 
Trading $\lambda_\mu$ for $q^a_\mu$ is not a change of variables, because the phases of $\lambda_\mu$ and $\eta$ are interlocked.
Nonetheless, with this extra hypothesis, $q^a$ can be recovered from this identification as a function of $\eta$ and $\underline\lambda$, $E = 4 ||\eta||^2 q $.

It would be therefore interesting to continue this analysis of spinorial action principles in the continuum to find those that correspond to specific spin-network functionals of particular interest, possibly along the lines of \cite{Wieland2014.07}. 
Another avenue of investigation should clarify the fate of shift and diffeomorphism symmetries in these continuous actions (e.g. diffeomorphisms are covered in \cite{Wieland2017c}).

\section{Haar intertwiner's dual in the face model \label{app_HaarRSOS}}

The simplest way to tackle the translation of the Haar inetertwiner to the face model representation, is to start all over again with the following rewriting of the solid torus amplitude (refer to the figure for details):
\begin{align}
\la &Z_{S\mathbb T_2} | \Psi^{\mathbb T_2}_{(j,\iota)} \ra = \notag\\
=& \Big[\prod_{l^*\notin R^*}\int \d h_{l^*} \Big]  \notag\\
&\Big[\prod_{l^*\in R^*} \int \d h^\text{t}_{l^*} \d h^\text{b}_{l^*} \d k^\text{t}_{l^*} \d k^\text{b}_{l^*} \Big] \notag\\
&\delta(K^\text{t}) \prod_{l^*\in R^*}\delta(H^\text{t}_{l^*} H^\text{b}_{l^*} ) \notag\\
&\prod_{l^*\in R^*}\delta(H_{l^*}^\text{t}) \delta(H_{l^*}^\text{b})  \prod_{f^*\atop f^*\cap R^* = \emptyset}\delta(H_{f^*}) \notag\\
& \Psi_{(j,\iota)}[h_{l^*\notin R^*}, h_{l^*\in R^*}= h_{l^*}^\text{t} h_{l^*}^\text{b} ]
\end{align}
This rewriting follows the idea that the solid torus has been cut open in a solid cylinder, i.e.
\begin{center}
\vspace{.5em}\includegraphics[scale=.2]{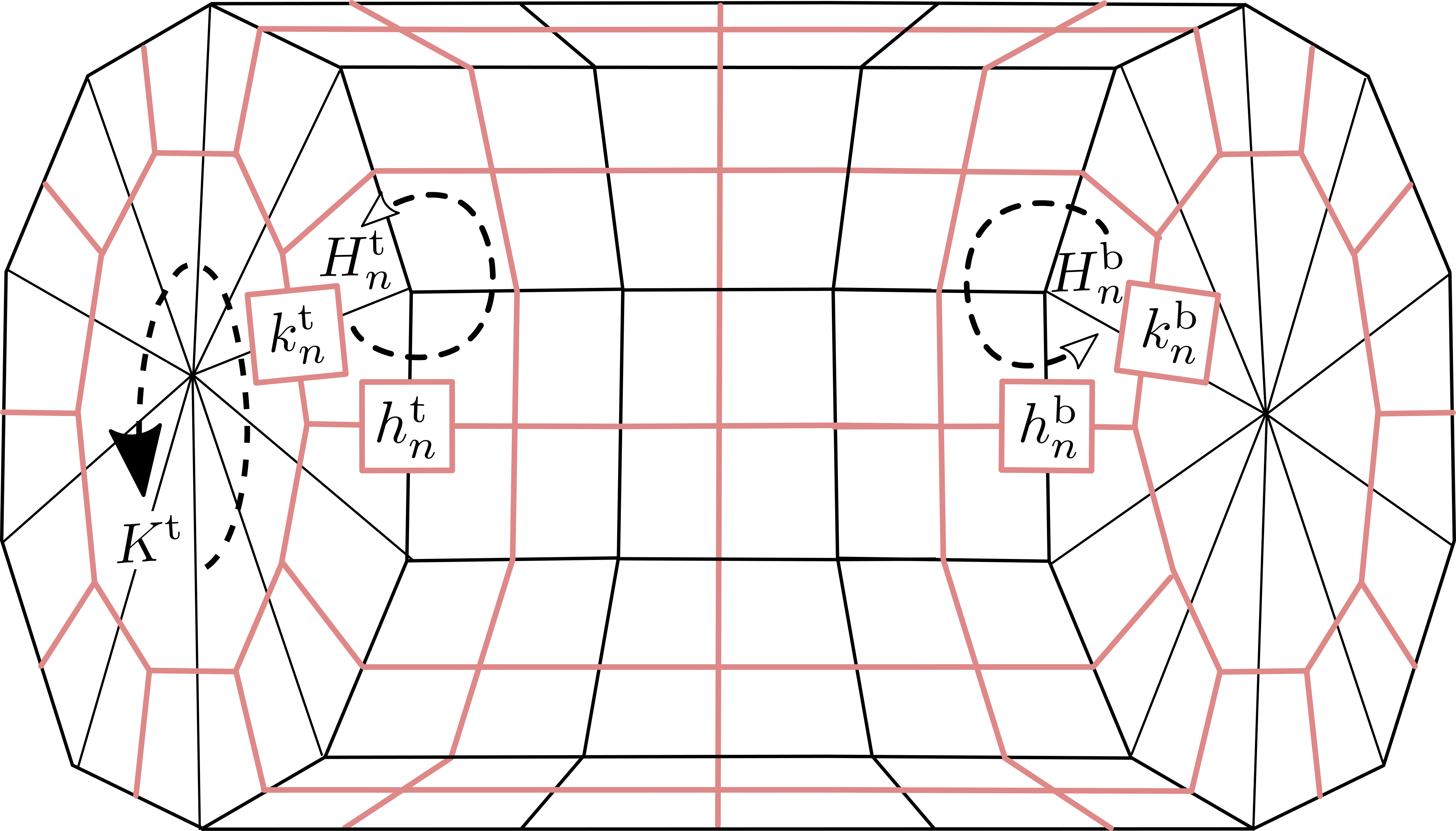}
\end{center}

Thus ($i$) the holonomies crossing the ring have been split in two parts, assoicated to the top and bottom basis of the cylinder (this split automatically implements the presence of a non-trivial longitudinal holonomy)
\be
h_{l^*\in R^*}= h_{l^*}^\text{t} h_{l^*}^\text{b};
\ee
also  ($ii$) new holonomy variables $k^\text{t,b}_{l^*}$ have been introduced which are dual the triangulation of the top and bottom bases of the solid cylinder (the labeling by $l^*$ is conventional); 
($iii$) finally, we see that a handful of new delta functions have introduced, their meaning is the following.

$\delta(K^\text{t})$ represent the flatness of the (new) top dual face,
\be
K^\text{t} = \overleftarrow\prod_{l^*\in R^*} k_{l^*},
\ee
and says that one of the two cycles of the torus is contractible (the analogue delta function for the bottom face would be redundant).

$\delta(H^\text{t}_{l^*} H^\text{b}_{l^*} )$ are the gluing conditions, where for the $n$-th dual edge $l^*\in R^*$ one has schematically
\be
H^\text{t}_{n} =   (h^\text{t}_{n+1})^{-1} k^\text{t}_n h^\text{t}_{n} H'_\text{t} 
\ee
with $H'_\text{t}$ representing the remaining holonomy around the top portion of the face cut in two by $R$---similarly for $H^\text{b}_{n}$. A twist can be implemented at this level, via a shifted delta $\delta(H^\text{t}_{n} H^\text{b}_{n+N_\gamma} )$. for simplicity will not purse this possibility here.

Finally, the delta functions on the second to last line simply represent the local flatness on the boundary of the cylinder. 

From the above expressions, one sees that all ``$h$'' holonomies appear three times as before---see Sec. \ref{sec_metricbc}. 
For what concerns the ``$k$'' holonomies, on the other hand, one sees that the $k^\text{t}$ also appear three times---once in $\delta(K^\text{t})$, once in $\delta(H^\text{t}_{l^*})$, and once in the gluing condition $\delta(H^\text{t}_{l^*} H^\text{b}_{l^*} )$---while the $k^\text{b}_{l^*}$ appear only twice---there is not $K^\text{b}$.

Expanding 
\be
\delta(K^\text{t}) = \sum_{J_{\text{c}}} d_{J_\text{c}} \chi^{J_\text{c}}(K^\text{t}) 
\ee
where the label `c' stands for `core', and momentarily forgetting about the $k^\text{b}$ variables, we are mathematically in the same situation we used to be in Sec. \ref{sec_metricbc}.
In fact, the solid cylinder is nothing but a sphere, and from this viewpoint $J_\text{c}$ is the distance of the vertex at the center of the top face from the center of the sphere.
From the viewpoint of the solid cylinder, however, $J_\text{c}$ represents the length of the core of the solid torus, which is summed over (with the weight above) because its conjugate variable, the holonomy around the opposite cycle, must be trivial.

Explicitly integrating out the $k^\text{b}_{l^*}$, which appear only twice each, essentially implements the gluing. 
This leaves us only  with variables appearing three times, which allows us to apply the mathematical procedure of Sec. \ref{sec_metricbc} which led to a face model.
The difference is now that faces across the gluing interact with a spin $J_\text{c}$.
The physical interpretation of this interaction from the face model perspective has still to be elucidated.

\bibliographystyle{bibstyle_aldo}
\bibliography{PRbdrydofs_biblio}

\end{document}